\documentclass[%
 onecolumn,
 amsmath,amssymb,
 aps,
prfluids,
]{revtex4-1}

\usepackage{float}\usepackage{graphicx}
\usepackage{subfig}
\usepackage{dcolumn}
\usepackage{bm}

\usepackage[dvipsnames]{xcolor}

\begin{document}

\preprint{APS/123-QED}

\title{Statistical transition to turbulence in plane channel flow}

\author{S\'ebastien Gom\'e}

 \email{sebastien.gome@espci.fr}
\author{Laurette S. Tuckerman}%
 \email{laurette.tuckerman@espci.fr}
\affiliation{%
Laboratoire de Physique et M\'ecanique des Milieux H\'et\'erog\`enes (PMMH), CNRS, ESPCI Paris, PSL Research
University, Sorbonne Universit\'e, Universit\'e Paris Diderot, Paris 75005, France
}%

\author{Dwight Barkley}
 \email{D.Barkley@warwick.ac.uk}

\affiliation{
 Mathematics Institute, University of Warwick, Coventry CV4 7AL, United Kingdom
}%

\date{\today}

\begin{abstract}
Intermittent turbulent-laminar patterns characterize the transition to turbulence in pipe, plane Couette and plane channel flows. 
The time evolution of turbulent-laminar bands in plane channel flow is studied via direct numerical simulations using the parallel pseudospectral code ChannelFlow in a narrow computational domain tilted by $24^{\circ}$ with respect to the streamwise direction. Mutual interactions between bands are studied through their propagation velocities. Energy profiles show that the flow surrounding isolated turbulent bands returns to the laminar base flow over large distances. Depending on the Reynolds number, a turbulent band can either decay to laminar flow or split into two bands. As with past studies of other wall-bounded shear flows, 
in most cases survival probabilities are found to be consistent with exponential distributions for both decay and splitting, indicating that the processes are memoryless. Statistically estimated mean lifetimes for decay and splitting are plotted as a function of the Reynolds number and lead to the estimation of a critical Reynolds number $Re_{\text{cross}}\simeq 965$, where decay and splitting lifetimes cross at greater than $10^6$ advective time units. The processes of splitting and decay are also examined through analysis of their Fourier spectra. The dynamics of large-scale spectral components seem to statistically follow the same pathway during the splitting of a turbulent band and may be considered as precursors of splitting.

\end{abstract}

\maketitle

\section{Introduction}

The route to turbulence in many wall-bounded shear flows involves intermittent laminar-turbulent patterns that evolve on vast space and time scales~(\cite{tuckerman2020patterns} and references therein). These states have received much attention over the years, both because of their intrinsic fascination and also because of their fundamental connection to critical phenomena associated with the onset of sustained turbulence in subcritical shear flows.  Below a critical Reynolds number, intermittent turbulence exists only transiently -- inevitably reverting to laminar flow, possibly after some very long time. Just above the critical Reynolds number, turbulence can become sustained in the form of intermittent laminar-turbulent patterns.

Flow geometry, specifically the number of unconstrained directions, plays an important role in these patterns. In flows with one unconstrained direction, large-scale turbulent-laminar intermittency can manifest itself only in that direction. Pipe flow is the classic example of such a system \cite{Reynolds:1883}, but other examples are variants such as duct flow \cite{takeishi2015localized} and annular pipe flow \cite{ishida2016transitional}, and also constrained Couette flow between circular cylinders where the height and gap are both much smaller than the circumference \cite{lemoult2016directed}. In terms of large-scale phenomena, these systems are viewed as one dimensional. Turbulent-laminar intermittency takes the comparatively simple form of localized turbulent patches, commonly referred to as puffs, interspersed within laminar flow \cite{darbyshire1995transition,nishi2008laminar,van2009flow}. In this case much progress has been made in understanding the localization of puffs and the critical phenomena associated with them \cite{hof2010eliminating,samanta2011experimental,avila2011onset,barkley2015rise,barkley2016theoretical,barkley2015rise}, including the scaling associated with one-dimensional directed percolation \cite{lemoult2016directed}.

In flow geometries with one confined and two extended directions, turbulent-laminar intermittency takes a more complex form that is dominated by turbulent bands which are oriented obliquely to the flow direction. Examples of such flows are Taylor-Couette flow \cite{coles1966progress,andereck1986flow,dong2009evidence,meseguer2009instability,Kanazawa_thesis,berghout2020direct,prigent2002large}, plane Couette flow \cite{prigent2002large,duguet2010formation}, plane channel flow \cite{tsukahara2014dns,brethouwer2012turbulent,fukudome2012large}, and a free-slip version of plane Couette flow called Waleffe flow \cite{waleffe1997self,chantry2016turbulent}. In terms of large-scale phenomena, one views these systems as two dimensional. Understanding the transition scenario in these systems is complicated by the increased richness of the phenomena they exhibit and also by the experimental and computational challenges involved in studying systems with two directions substantially larger than the wall separation. So large are the required dimensions that only for a truncated model of Waleffe flow has it thus far been possible to verify that the transition to turbulence is of the universality class of two-dimensional directed percolation \cite{chantry_universal}.

Between the one-dimensional and fully two-dimensional cases are the numerically obtainable restrictions of planar flows to long, but narrow, periodic domains tilted with respect to the flow direction \cite{barkley2005computational}. These  domains restrict turbulent bands to a specified angle. They have only one long spatial direction, thereby limiting the allowed large-scale variation to one dimension, but they permit flow in the narrow (band-parallel) direction, flow that is necessary for supporting turbulent bands in planar shear flows. Such computational domains were originally proposed as minimal computational units to capture and understand the oblique turbulent bands observed in planar flows \cite{barkley2005computational}. Tilted computational domains have subsequently been used in numerous studies of transitional wall-bounded flows, notably plane Couette flow\cite{barkley2007mean,tuckerman2011patterns,shi,lemoult2016directed,reetz2018invariant} and plane channel flow \cite{tuckerman2014turbulent, paranjape2020oblique}. Lemoult \emph{et al.} \cite{lemoult2016directed} showed that in tilted domains plane Couette flow exhibits a transition to sustained turbulence in the directed percolation universality class. Reetz, Kreilos \& Schneider \cite{reetz2018invariant} computed a state resembling a periodic turbulent band in plane Couette flow while Paranjape, Duguet \& Hof \cite{paranjape2020oblique} computed localized traveling waves in plane channel flow as a function of the Reynolds number and the tilt angle.
Shi, Avila \& Hof \cite{shi} used simulations in a tilted domain to measure decay and splitting lifetimes in plane Couette flow 
and it is this approach that we apply here to plane channel flow. 

We mention two important points concerning the relevance of turbulent bands in narrow tilted domains to those in plane channel flow in large domains. The first is that a regime in transitional channel flow has been discovered at Reynolds numbers lower than those studied here in which turbulent bands elongate at their downstream end while they retract from their upstream end \cite{xiong2015turbulent,Kanazawa_thesis,tao2018extended,xiao2020growth,ShimizuPRF2019}. 
Such bands of long but finite length are excluded in narrow tilted domains.
In full two-dimensional domains and at lower Reynolds numbers, this one-sided regime takes precedence over the transition processes that we will describe here. The second point is that critical Reynolds numbers obtained in narrow tilted domains \cite{shi,Chantry_private} have been found to agree closely with transition thresholds found in the full planar setting \cite{bottin1998statistical, bottin1998discontinuous, duguet2010formation,chantry_universal} in both plane Couette flow and in stress-free Waleffe flow. We will return to both of these points in Sec.~\ref{sec:Discussion}.

Here we study the onset of turbulent channel flow in narrow tilted domains. We follow closely the work of Shi, Avila \& Hof \cite{shi} on plane Couette flow. We are particularly focused on establishing the time scales and Reynolds numbers associated with the splitting and decay processes.

\section{Numerical procedure and choice of dimensions}

Plane channel flow is generated by imposing a mean or bulk velocity $U_\text{bulk}$ on flow between two parallel rigid plates. The length scales are nondimensionalized by the half-gap $h$ between the plates. 
Authors differ on the choice of velocity scales for nondimensionalizing channel flow, but one standard choice, that we adopt here, is to use 
$3 U\text{bulk} /2$. This is equal to the centerline velocity $U_\text{cl}$ of the 
corresponding laminar parabolic flow since
\begin{equation}
U_\text{bulk}=\frac{1}{2}\int_{-1}^{+1}U_\text{cl}(1-y^2) dy = \frac{2}{3}U_\text{cl}
    \end{equation}
The Reynolds number is then defined to be $Re= U_\text{cl} h/\nu = 3 U_\text{bulk} h/(2\nu)$.

\begin{figure}
    \centering
    \includegraphics[width=\columnwidth]{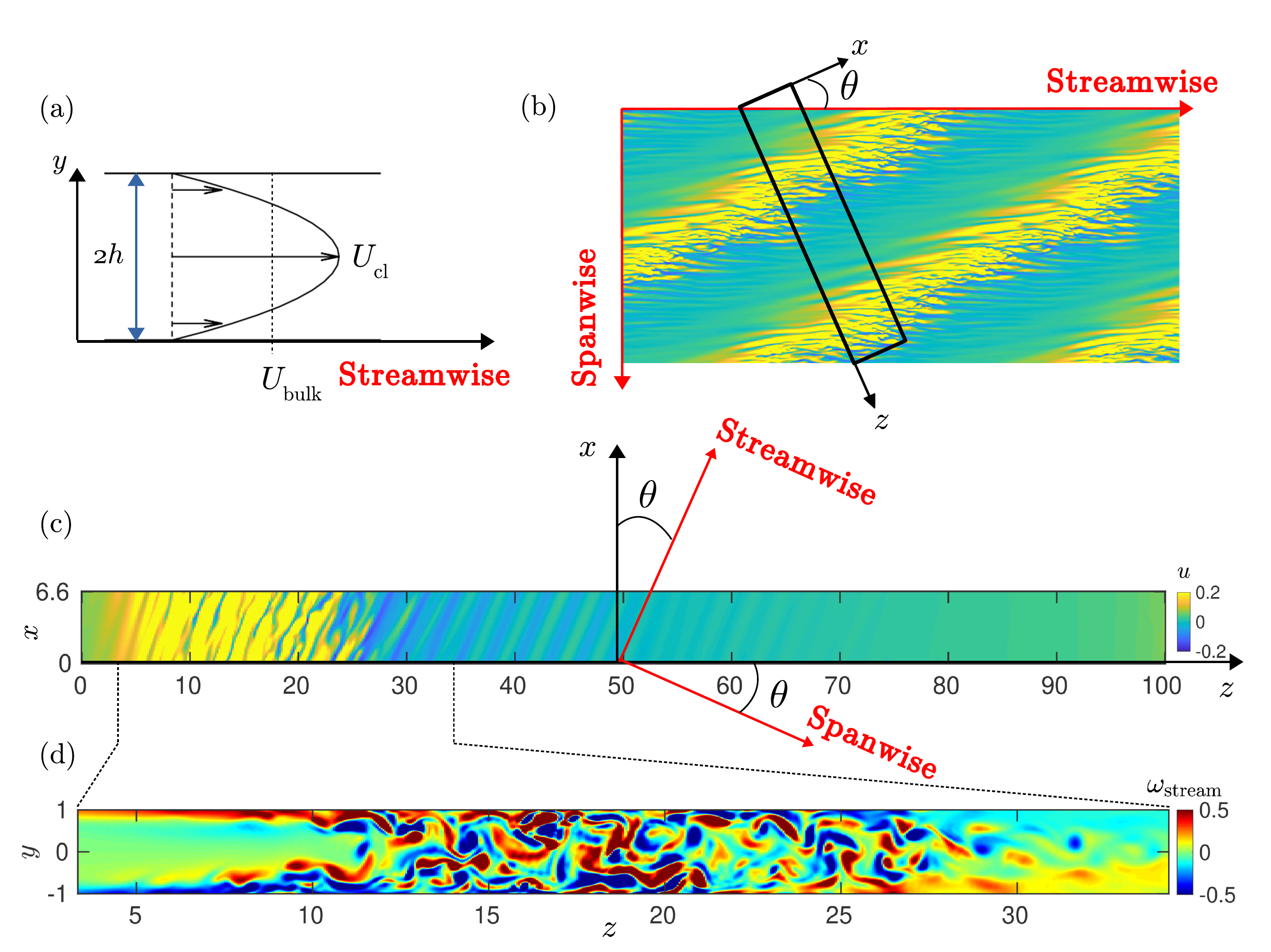}
    \caption{(a) Sketch of the laminar profile. (b) Visualization of turbulent bands in a $240\times 108$ streamwise-spanwise domain at $Re=1000$. Colors indicate the streamwise velocity in the $y=-0.8$ plane. A superimposed black box illustrates a long-narrow computational domain, tilted with an angle $\theta$ relative to the streamwise direction. (c) and (d) Structure of a turbulent-laminar pattern computed in a tilted domain at $Re=1200$. Plot (c) shows the $x$ component of the velocity in the $(x,z)$ plane at $y=-0.8$. The streamwise and spanwise directions are indicated in red. Plot (d) shows streamwise vorticity in a $(y,z)$ plane with the vertical $y$ scale stretched by a factor of 2. Only the portion of the computational domain containing the turbulent region is shown in (d). As seen in (c), on the downstream side of the turbulent region the flow exhibits weak straight streaks, oriented in the streamwise direction, that slowly diminish as the flow returns laminar.}
   
    \label{fig:tilted_box}
\end{figure}

The computational domain used in this study is tilted with respect to the streamwise direction, as illustrated in Fig. \ref{fig:tilted_box}(b). Its wall-parallel projection is a narrow doubly-periodic rectangle with the narrow dimension (labelled by the $x$ coordinate) aligned along the turbulent band. The long dimension of the domain (labelled by the $z$ coordinate) is orthogonal to the bands, i.e. it is aligned with the pattern wavevector. The relationship between streamwise-spanwise coordinates and $(x,z)$ coordinates is:
\begin{subequations}
\label{tilted}
\begin{eqnarray}
\mathbf{e_{\text{streamwise}}} =& & \cos{\theta} \, \mathbf{e}_x + \sin{\theta} \, \mathbf{e}_z \\
\mathbf{e_{\text{spanwise}}} = & - & \sin{\theta } \, \mathbf{e}_x + \cos{\theta} \, \mathbf{e}_z 
\quad 
\end{eqnarray}
\end{subequations}
The wall-normal coordinate is denoted $y$ and is independent of the tilt. 

The angle in this study is fixed at $\theta=24^{\circ}$, as has been used extensively in the past. The tilt angle of the domain imposes a fixed angle on turbulent bands. (Turbulent bands at larger angles have also been observed in large or tilted domains.) The narrowness of the computational domain in the $x$ direction prohibits any large-scale variation along turbulent bands, effectively simulating infinitely long bands. These restrictions of a tilted domain have both advantages and disadvantages for simulations of transitional turbulence. We return to this in the discussion. 
 
We have carried out direct numerical simulations (DNS) using the parallelized pseudospectral C++-code  ChannelFlow \cite{channeflow}. This code simulates the incompressible Navier-Stokes equations in a periodic channel by employing a Fourier-Chebychev spatial discretization, fourth-order semi-implicit backwards-differentiation time stepping, and an influence matrix method with Chebyshev tau correction to impose incompressibility in the primitive-variable formulation. 
The velocity field is decomposed into a parabolic base flow and a deviation, $\mathbf{U} = \mathbf{U_{\text{base}}} + \mathbf{u}$, where the deviation field $\mathbf{u}$ has zero flux. Simulating in the tilted domain gives velocity components $\mathbf{u}= (u,v,w)$ aligned with the oblique coordinates $(x,y,z)$.
All kinetic energies reported here are those of the deviation from laminar flow $\frac{1}{2}\int (u^2+v^2+w^2)$,
rather than the turbulent kinetic energy (defined to be that of the deviation from the mean velocity).

Most of the simulations presented have been carried out in a domain with dimensions ($L_x,L_y,L_z) = (6.6,2,100$). The numerical resolution is $(N_x,N_y,N_z)=(84,64,1250)$, which both ensures that $\Delta x = \Delta z \simeq 0.08$ and that $\Delta y$ varies from $\Delta y = \cos (31\pi /64) = 0.05$ at $y=0$ to $\Delta y = 1-\cos (\pi/64) = 0.001$ at $y=\pm 1$. This resolution has been shown to be sufficient to simulate small turbulent scales at low Reynolds numbers (Kim \emph{et al.} \cite{kim1987turbulence}, Tsukahara \emph{et al.} for $Re=1370$ \cite{tsukahara2014dns}).

In the Fourier-Chebychev discretization the deviation velocity is expressed as:
\begin{equation}
    \mathbf{u} = \sum_{-N_x/2+1}^{N_x/2} \sum_{-N_z/2+1}^{N_z/2} \sum_{0}^{N_y} \mathbf{\hat{u}}_{m_x,m_y,m_z} e^{i (k_x m_x x + k_z m_z z)}  {T}_{m_y} (y)
\label{eq:specsum}\end{equation}
\noindent where $k_x=2\pi/L_x$, $k_z=2\pi/L_z$, $\mathbf{\hat{u}}_{m_x,m_y,m_z} $ are the Fourier-Chebyshev coefficients, and ${T}_{m_y} (y)$ are the Chebychev polynomials. For brevity, we will refer to $m_x$ and $m_z$ (rather than $m_x k_x$, $m_z k_z$) as wavenumbers.

The structure of a typical turbulent band in this domain is shown on Fig. \ref{fig:tilted_box}. A series of straight periodic streaks is visible downstream of the turbulent band, whereas the upstream laminar-turbulent interface is much sharper. Streaks are visible here as streamwise velocity modulated along the spanwise direction. They are wavy in the core of the turbulent zone, in accordance with the self-sustaining process of transitional turbulence \cite{waleffe1997self}.

Our choice for the standard domain dimensions, ($L_x,L_y,L_z) = (6.6,2,100$), is dictated as follows: $L_y=2$ is fixed by non-dimensionalization.  The choice of the short dimension $L_x$ is dictated by the natural streak wavenumber. In plane Couette flow, this was found to be approximately $L_{x, \text{Couette}}=10 = 4/\sin{24^{\circ}}$ \cite{hamilton1995regeneration}, and widely used since \cite{barkley2005computational, shi}. Chantry \emph{et al.} showed that the correspondence between length scales in plane Couette and plane channel flows is $h_\text{Poiseuille} \simeq 1.5 h_{\text{Couette}}$ (by doubling the Couette height and subtracting the resulting spurious mid-gap boundary layer \cite{chantry2016turbulent}). This leads to an optimal short dimension in a $24^{\circ}$ box of $L_{x, \text{Poiseuille}} = 6.6$. ($L_x=6.6$ has also been used in \cite{paranjape2020oblique}, whereas $L_x=10$ was used in \cite{tuckerman2014turbulent}.) $L_z = 100$ is chosen to be sufficiently large that periodicity in the $z$-direction does not have a significant effect on the turbulent band dynamics, as we will see in the next section. 

\section{Band velocity and interaction length}
\label{sec:band_velocity}

As in pipe flow \cite{hof2010eliminating,samanta2011experimental,barkley2016theoretical}, bands in channel flow interact when sufficiently close and this can affect the quantities we seek to measure. For example, in a one-dimensional directed percolation model \cite[p. 167]{shih2017spatial}, the time scales observed for decay and splitting increase strongly with the inter-band distance, while the critical point increases weakly.
We wish to choose the length $L_z$ of our domain to be the minimal distance above which bands can be considered to be isolated. 

Unlike their counterparts in plane Couette flow, turbulent bands in plane channel flow are not stationary relative to the bulk velocity $U_\text{bulk}$. As in pipe flow \cite{avila2011onset,barkley2015rise}, bands move either faster or slower than the bulk velocity, depending on the Reynolds number \cite{tuckerman2014turbulent}.
One important way in which the interaction between bands manifests itself is by a change in propagation speed. 
\begin{figure}
        \centering
        \subfloat[]{\includegraphics[width=0.39 \columnwidth]{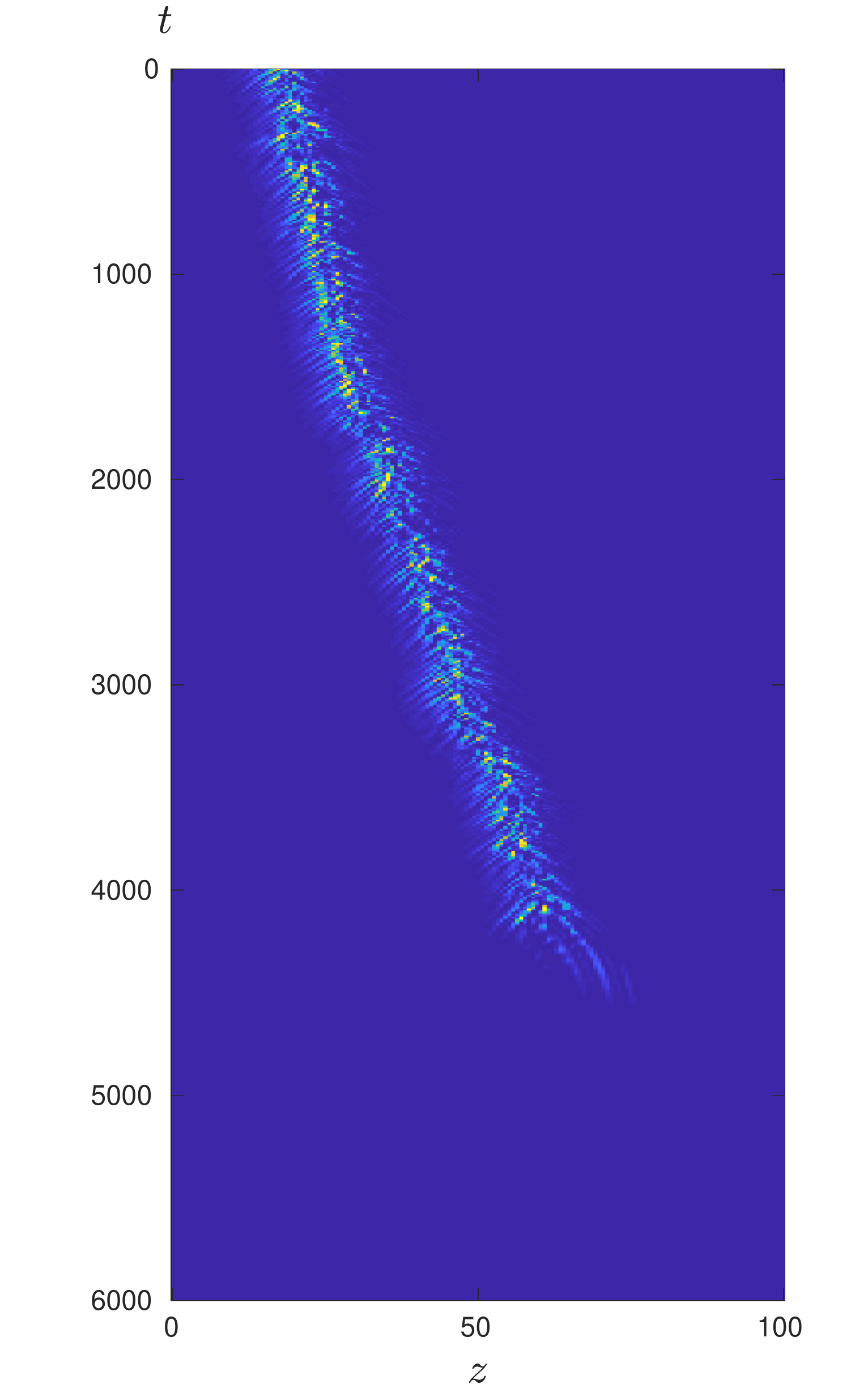}
        \label{fig:Lz100_R830}}
        \subfloat[]{\includegraphics[width=0.39 \columnwidth]{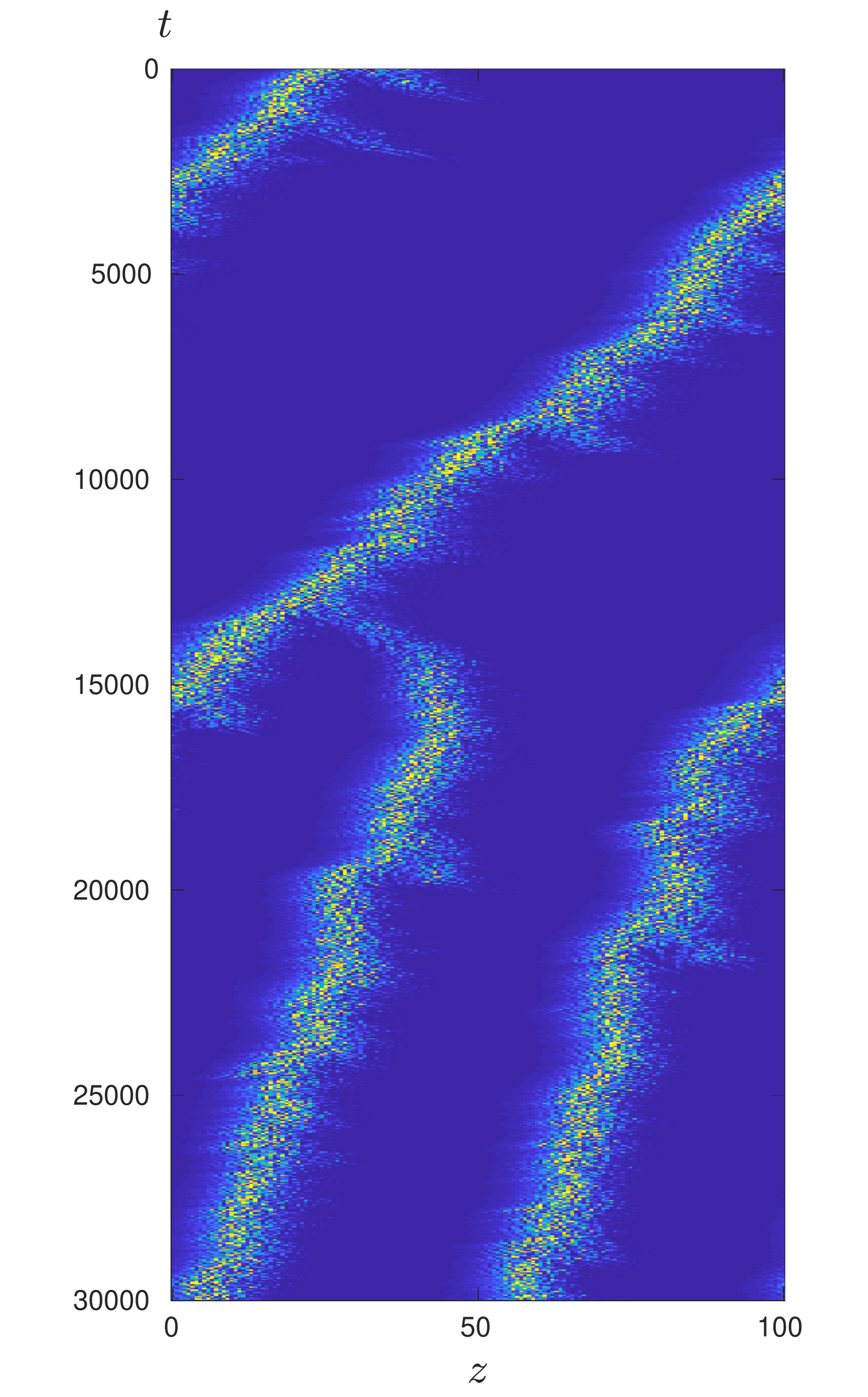}
        \label{fig:Lz100_R1100}}     \subfloat[]{\includegraphics[width=0.195 \columnwidth]{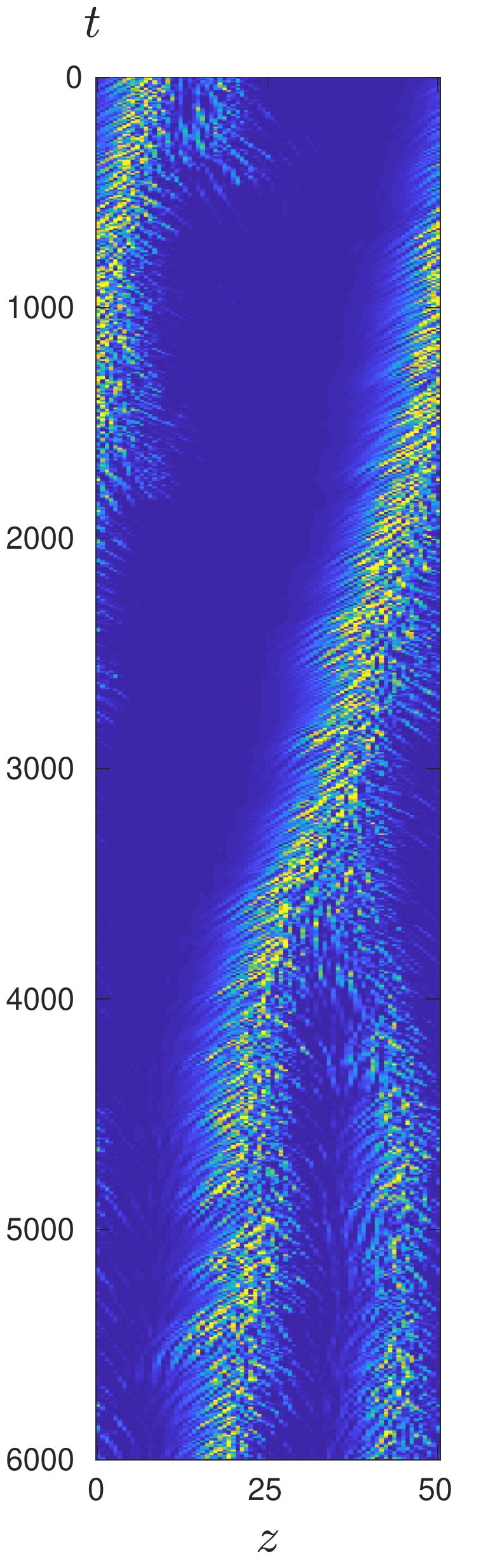}
        \label{fig:Lz50_R1200}}
        \caption{Space-time diagrams of turbulent bands in a frame moving at the bulk velocity, $U_{\rm bulk}$, with (a) $Re=830$, $L_z=100$, (b) $Re=1100$, $L_z=100$, (c) $Re=1200$, $L_z=50$. Colors show the perturbation energy $E=\frac{1}{2}(u^2 + v^2 + w^2)$ as a function of $z$ and $t$, sampled in the $y=-0.8$ plane at a arbitrary value of $x$ (yellow: $E=0.1$, blue: $E=0$). Average band propagation velocities, relative to $U_{\rm bulk}$, and the degree of fluctuations can be discerned from diagrams. Case (a) is an example of a band moving downstream relative to $U_\text{bulk}$, which occurs for $Re\lesssim 1000$, and then decaying. In case (b), a single band in a domain with $L_z=100$ splits into two bands, resulting in a pair of bands separated in $z$ by distance 50 = $L_z/2$. The change in velocity resulting from a decrease in interaction distance is evident. Note, however, that the time range covered in the plot is large, which visually accentuates the effect. Case (c) shows band splitting in a domain of size $L_z=50$. The resulting bands are closely spaced and interact strongly.}
    \label{fig:speed_time}
\end{figure}

Figure~\ref{fig:speed_time} illustrates some of the key issues via spatio-temporal plots of turbulent bands in a reference frame moving at the bulk velocity. Note that the imposition of periodic boundary conditions in $z$ leads to interaction across the boundary. Figure \ref{fig:Lz100_R830} illustrates a typical long-lived turbulent band at $Re \lesssim 1000$. The band moves slowly in the positive $z$ direction, \emph{i.e} downstream relative to the bulk velocity, and then decays, i.e. the flow relaminarizes.
 
Figure \ref{fig:Lz100_R1100} illustrates a typical band splitting at $Re=1100$, for which bands move upstream relative to the bulk velocity. At $t\simeq 13\,000$ a daughter band emerges from the downstream side of the parent band, very much like puff splitting observed in pipe flow~\cite{avila2011onset,shimizu2014splitting}. Following the split, the distance between bands decreases (from $L_z=100$ to $L_z/2=50$), thereby increasing the band interaction, as can be seen by a change in the propagation velocity following the split. The time range in Fig.~\ref{fig:Lz100_R1100} is very long and this visually accentuates the speed change. The absolute speed change following the split is approximately 1\% of the bulk velocity.
Figure \ref{fig:Lz50_R1200} presents a band splitting in a box of size $L_z=50$ at $Re=1200$ and shows a more pronounced difference in propagation velocities between the single band and its two offspring. The quasi-laminar gap separating the two offspring bands is quite narrow and hence the bands can be assumed to strongly interact.
The spatio-temporal diagrams of Fig.~\ref{fig:speed_time} also show that the size of turbulent bands increases slightly with $Re$, and moreover that fluctuations in the size and propagation speed become greater. Fluctuations are more pronounced on the downstream side of bands. 
\begin{figure}
    \centering
    \includegraphics[width=0.6\columnwidth]{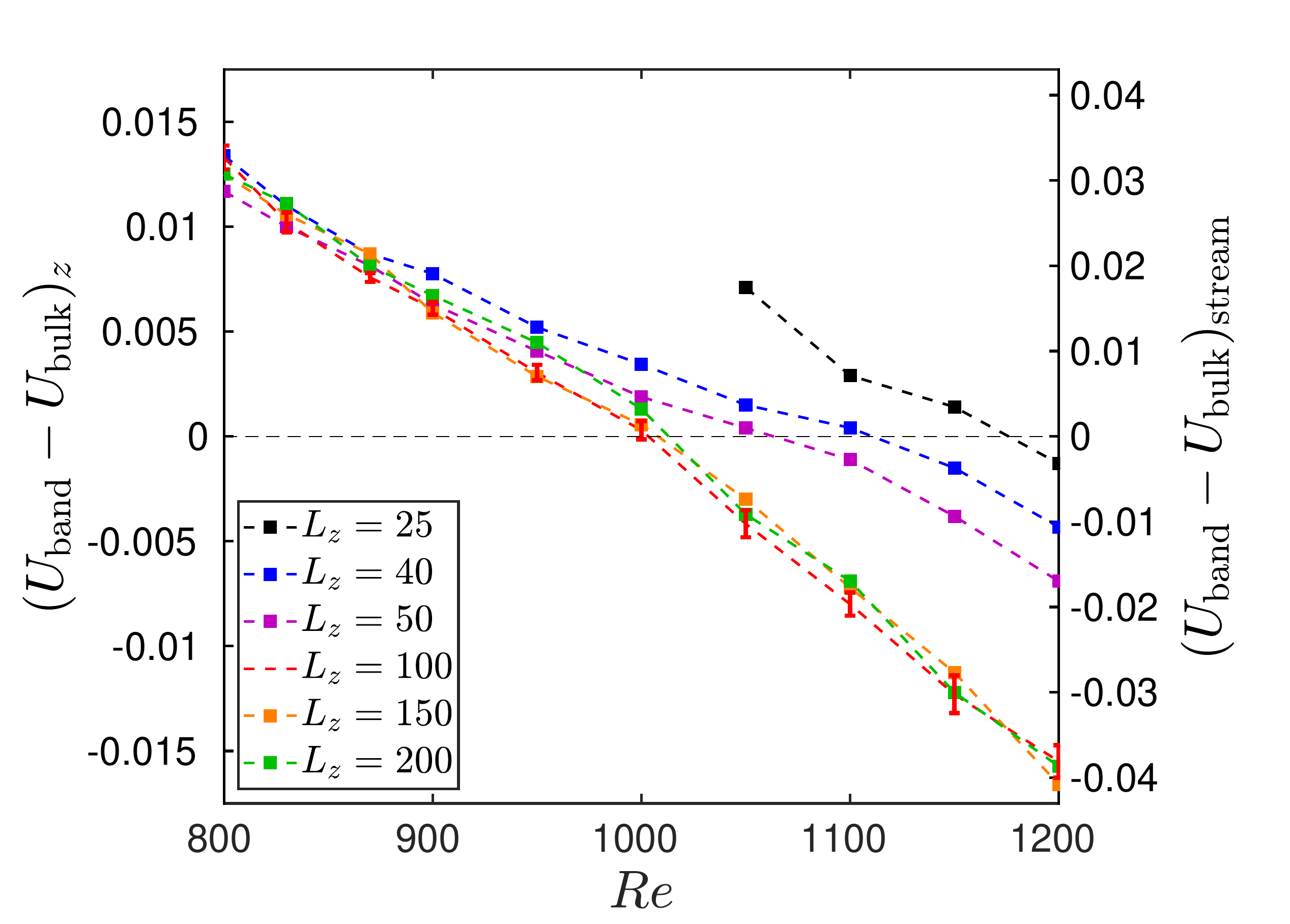}
    \caption{Dependence of the band propagation velocity on the Reynolds number and on the inter-band distance $L_z$ (left axis: $z$ velocity, right axis: streamwise velocity). Normal-approximated error bars are shown for $L_z=100$.}
    \label{fig:speed}
\end{figure}
\begin{figure}
        \centering
       \subfloat[]{\includegraphics[width=0.49\textwidth]{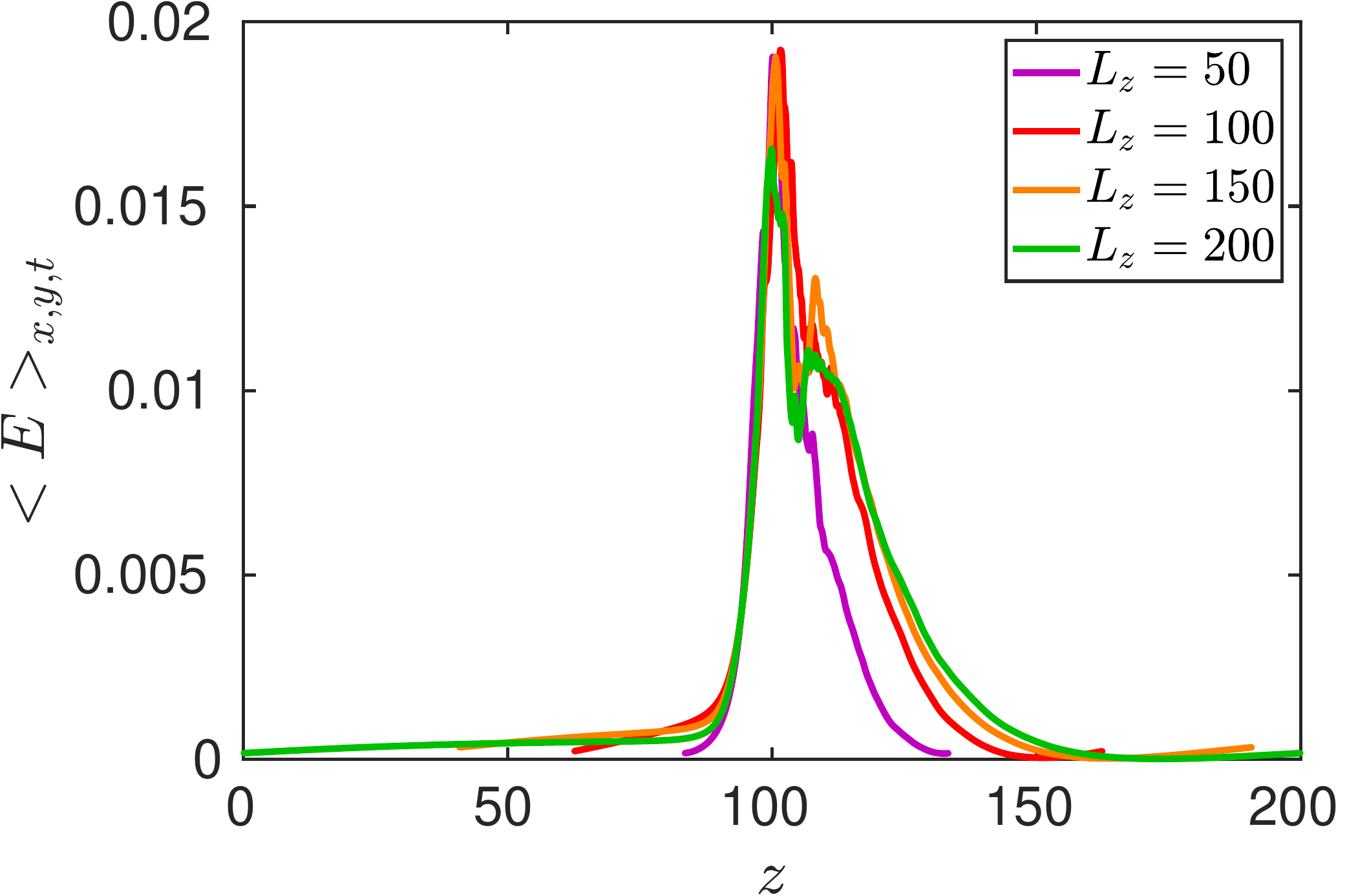}  \label{fig:Exyt_z_lin}}
        \hspace*{-0.5em}
        \subfloat[]{\includegraphics[width=0.49\textwidth]{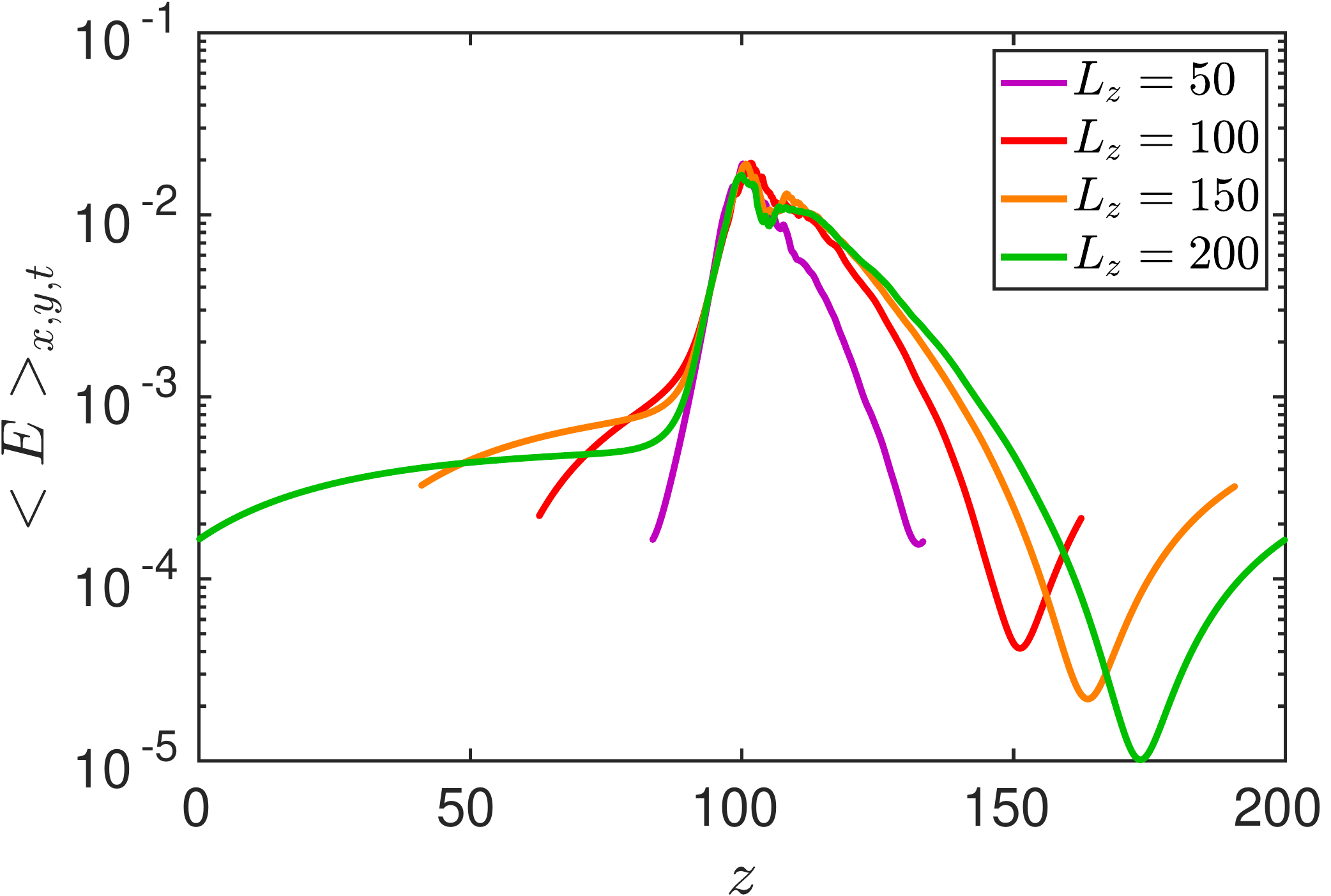}
        \label{fig:Exyt_z_log}}
    \caption{Energy averaged over $x$, $y$ and $t$ as a function of $z$, for different $L_z$, in (a) linear and (b) logarithmic scales, for a one-band state at $Re=1000$}
    \label{fig:Exyt_z}
\end{figure}
More quantitatively, we have measured the propagation speed, $U_{\text{band}}$, of single turbulent bands over a range of $Re$ in domains of different lengths $L_z$, as shown in Fig. \ref{fig:speed}. Periodic boundary conditions in $z$ set the center-to-center interaction distance between bands to the domain length $L_z$. 
Single bands were simulated for up to a total of 70000 time units. Error bars (only shown in case $L_z=100$ for clarity) represent normal-approximated confidence intervals for time-weighted velocity measurements over the multiple simulations comprising the total simulation time. 
Care was taken to discard pushing effects due to missed splittings or decays that may deviate the band from its average velocity. An initial time $t_0>0$ was subtracted to eliminate the effect of the initial conditions (see Sec.~\ref{sec:analysis} and \ref{sec:stochastic}).


We find that the band speed becomes independent of $L_z$ for $L_z \gtrsim 100$. The speeds vary approximately linearly with $Re$, over the range studied, and remain close to the bulk velocity: $|U_\text{band}-U_\text{bulk}|$ is less than 2\% of $U_\text{bulk}$.  For values of $L_z < 100$, speeds are shifted upwards, and their slopes vary from the slope at higher $L_z$. Note that bands at $L_z=25$ are not sustained for $Re \lesssim 1050$. Values at $L_z=40$ are similar to those reported in a domain of the same size in \cite{tuckerman2014turbulent}; Figure \ref{fig:speed} shows that this inter-band separation is too small to be in the asymptotic regime. (In addition, here the streamwise velocity is defined as $v_z/\sin\theta$, i.e. such that its projection in the $z$ direction is the $z$ velocity, whereas in \cite{tuckerman2014turbulent} it is defined to be $v_z\sin\theta$, i.e. the projection of the $z$ velocity along the streamwise direction.)

The streamwise band speeds observed here compare with what is known for puff speeds in pipe flow. For Reynolds numbers near where the puff speed equals the bulk velocity, the speed is given by $U_\text{p}-\bar U \simeq -2.4 \times 10^{-4} (Re - 1995)$, where $U_\text{p}$ is the nondimensional puff speed and $\bar U = 1$ is the nondimensional bulk velocity for pipe flow. (This expression comes from the data given in supplemental material for Ref.~\cite{avila2011onset}.) Making a linear approximation to the data in Fig.~\ref{fig:speed}, the streamwise band speeds can be approximated by $(U_\text{band}-U_\text{bulk})_\text{stream} \simeq -1.7 \times 10^{-4} (Re - 1000).$  Thus we find that variation of speed with Reynolds number is of the same magnitude in the two cases, that is the coefficients $-2.4 \times 10^{-4}$ and $-1.7 \times 10^{-4}$ are comparable. Both coefficients are negative reflecting that the downstream speed decreases as Reynolds number increases. (The reason for this is discussed at length for pipe flow in \cite{barkley2015rise,barkley2016theoretical}.) If one uses $2h$ for the length scale and bulk velocity for the velocity scale in channel flow, the coefficient for channel flow changes slightly to become $-1.9 \times 10^{-4}$. Detailed comparisons beyond this are not obviously meaningful without a precise way to map the Reynolds numbers between the two flows.

We also compare the kinetic energy profile in $z$ of stationary single bands at $Re=1000$, calculated in domains with $L_z$ between 50 and 200. Figure \ref{fig:Exyt_z_lin} shows the kinetic energy, i.e. the deviation from laminar flow, averaged over $x$, $y$, and $\Delta T=1000$, as a function of $z$, centered at $z=100$. We see a strong peak and width that, except for $L_z=50$, are nearly independent of $L_z$. The logarithmic representation of Fig.~\ref{fig:Exyt_z_log} highlights the weak tails of the turbulent bands. Except for $L_z=50$, all have an upstream "shoulder", i.e. a change in curvature followed by a plateau. All have a downstream minimum, whose position depends on $L_z$: for $L_z=50$ and 100, it is located halfway from the peak to its periodic repetition; for $L_z>100$ the ratio of this distance to $L_z$ decreases with increasing $L_z$. 
We doubled the resolution in the $z$ direction, and observed very little effect ($<2\%$) on the localization of the minimum.

Localized turbulent regions have been studied in other realizations of wall-bounded shear flows.
For exact computed solutions of plane channel flow,
the downstream spatial decay is observed to be more rapid than the upstream decay
\cite{zammert2014streamwise,zammert2016streamwise,paranjape2020oblique}, as in our case. In plane Couette flow \cite{barkley2005computational,brand2014doubly}, the upstream and downstream spatial decay rates are equal, by virtue of symmetry, while those of pipe flow show a strong dependence of the upstream decay rate on Reynolds number \cite{ritter2018analysis}. Asymmetry between upstream and downstream spatial decay rates is also seen in turbulent spots in boundary layer flow \cite{marxen2019turbulence} and in Poiseuille-Couette flow \cite{klotz2017Couette}.

Notwithstanding the long-range weak tails in Fig.~\ref{fig:Exyt_z_log}, we believe that turbulent bands in domains of at least $L_z=100$ can be considered as isolated: the quasi-laminar gap is sufficiently wide that one band does not substantially affect its neighbor and modify its velocity.

\section{Analysis of decay and splitting}
\label{sec:analysis}

\begin{figure}
        \centering
        \subfloat{\includegraphics[width=0.8\textwidth]{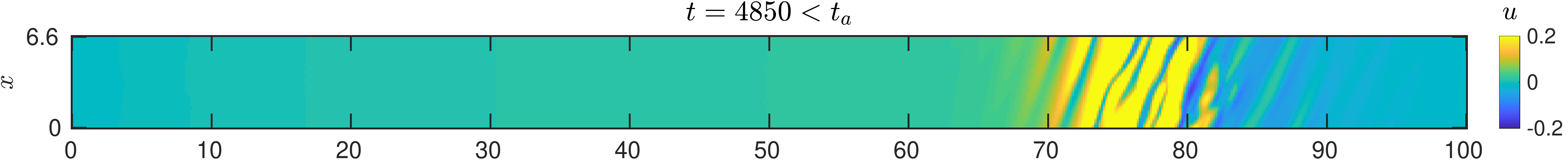}}\\
        \vspace*{-0.8em}
        \subfloat{\includegraphics[width=0.8\textwidth]{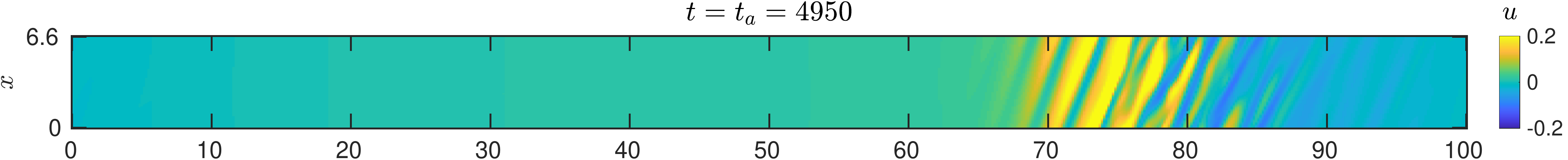}}\\       
        \vspace*{-0.8em}
        \subfloat{\includegraphics[width=0.8\textwidth]{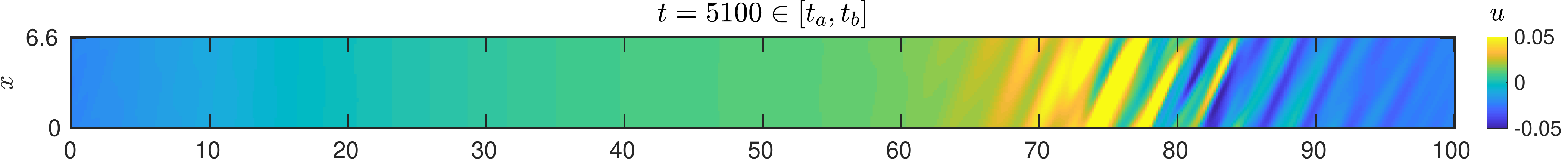}}\\        
        \vspace*{-0.8em}
        \subfloat{\includegraphics[width=0.8\textwidth]{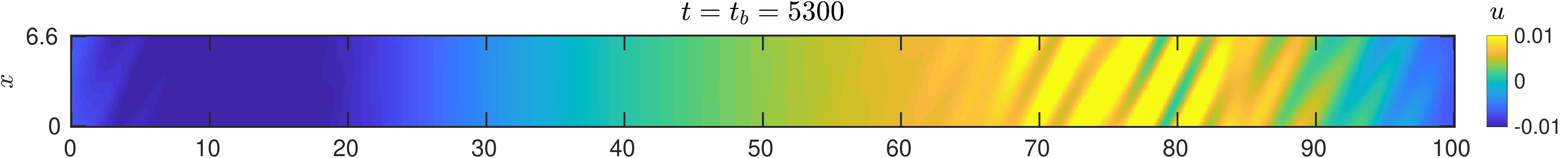}}\\
        \vspace*{-0.8em}
        \subfloat{\includegraphics[width=0.8\textwidth]{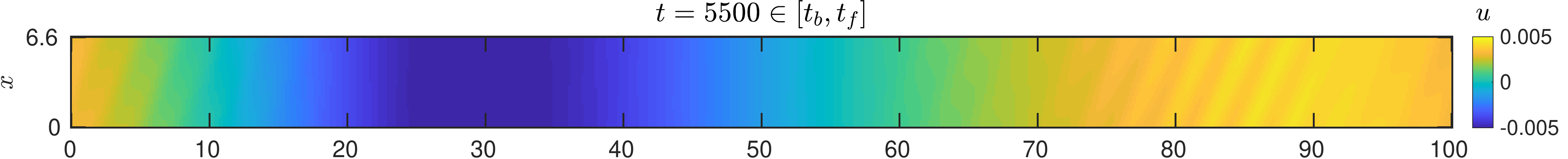}}\\
        \vspace*{-0.8em}
        \subfloat{\includegraphics[width=0.8\textwidth]{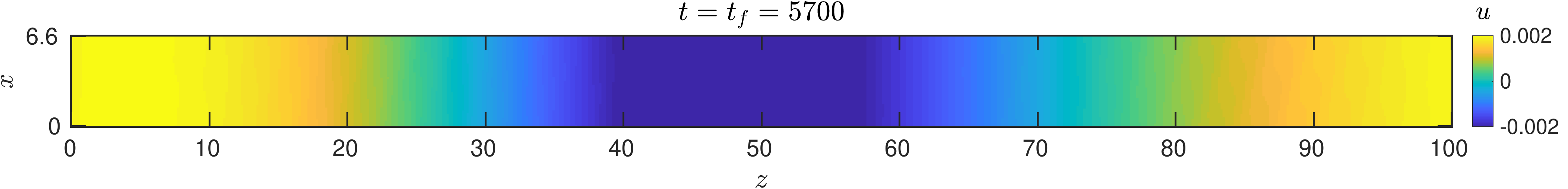}}\\
    \caption{Band decay at $Re=830$. Plotted is the $x$ velocity in $(x,y)$ planes at $y=-0.8$. For clarity the color scale changes over time.}
    \label{fig:decay_slices}
\end{figure}

\begin{figure}
    \centering
    \includegraphics[width=0.6\textwidth]{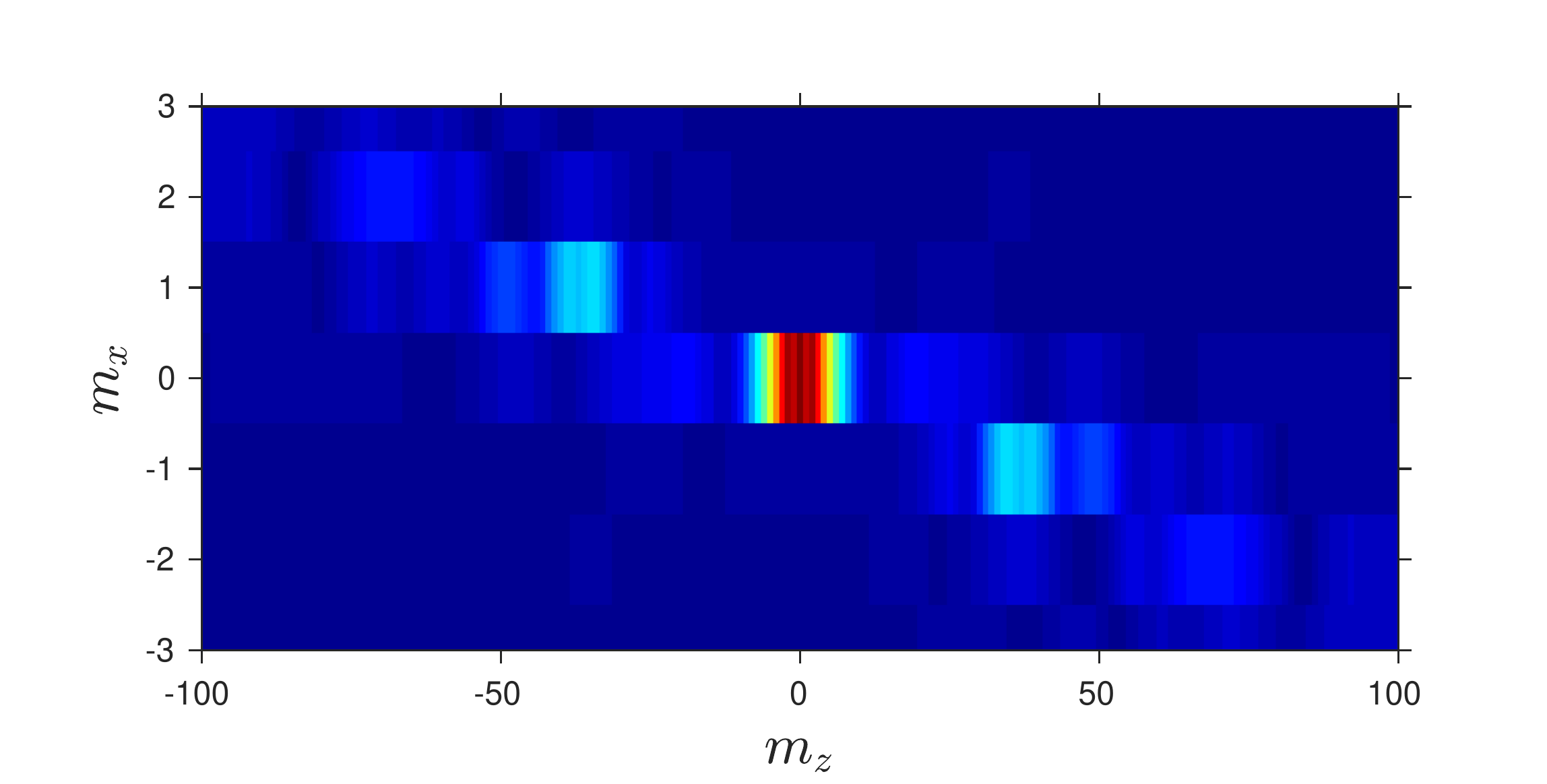}
    \caption{Example of a $(x,z)$ Fourier spectrum of the $x$ velocity $u$ in the $y=-0.8$ plane, for a turbulent band at $Re=830$. Colors show the modulus of spectral coefficients, spanning from 0 (blue) to 0.02 (red). The modulus of components $(m_x,-m_z)$ and $(-m_x,m_z)$ are equal since the velocity is real.}
    \label{fig:2D_spec}
\end{figure}

\begin{figure}
\centering
     \subfloat[$t_a=4950$]{\includegraphics[width=0.51\textwidth]{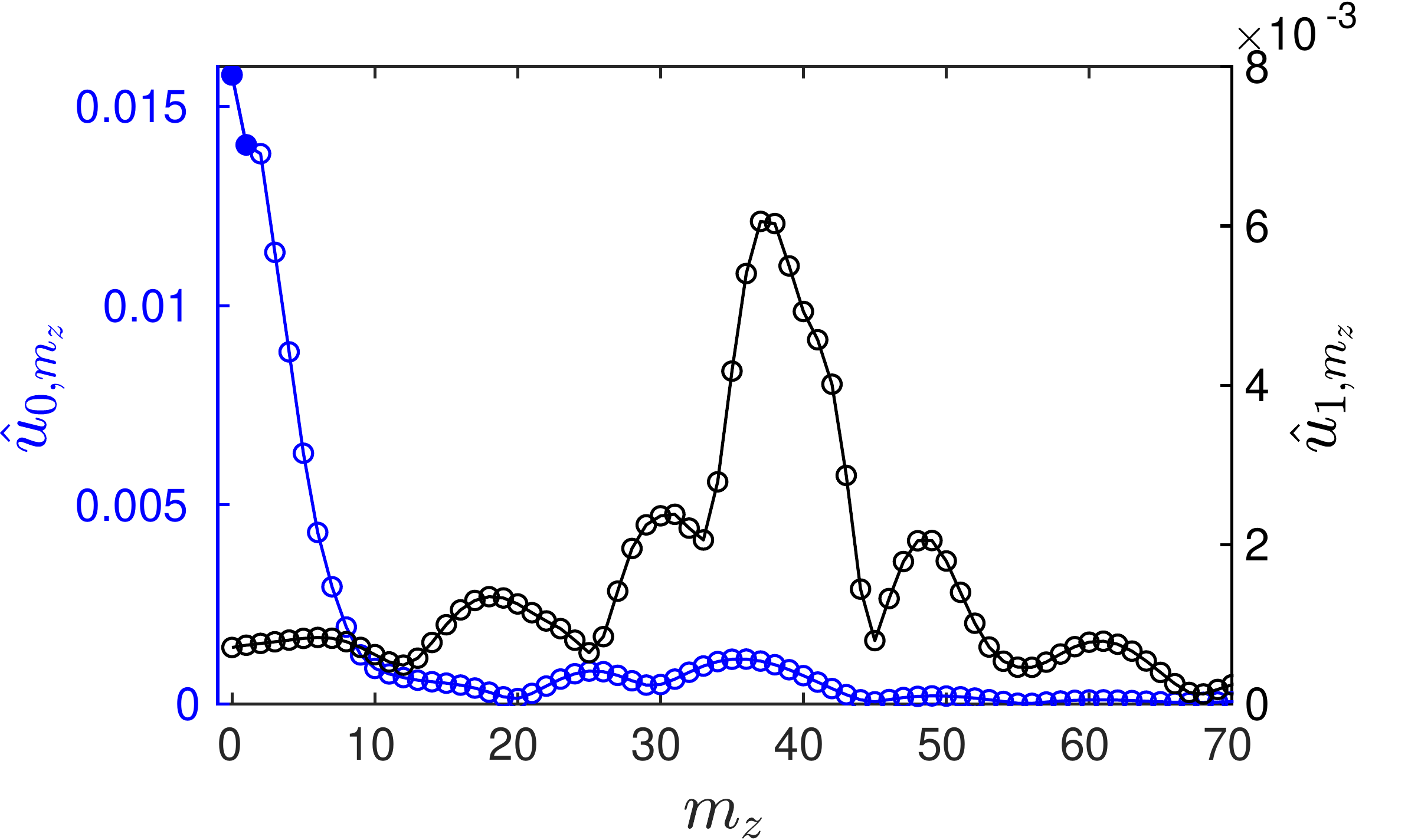}}
        \subfloat[$t_f=5700$]{\includegraphics[width=0.49\textwidth]{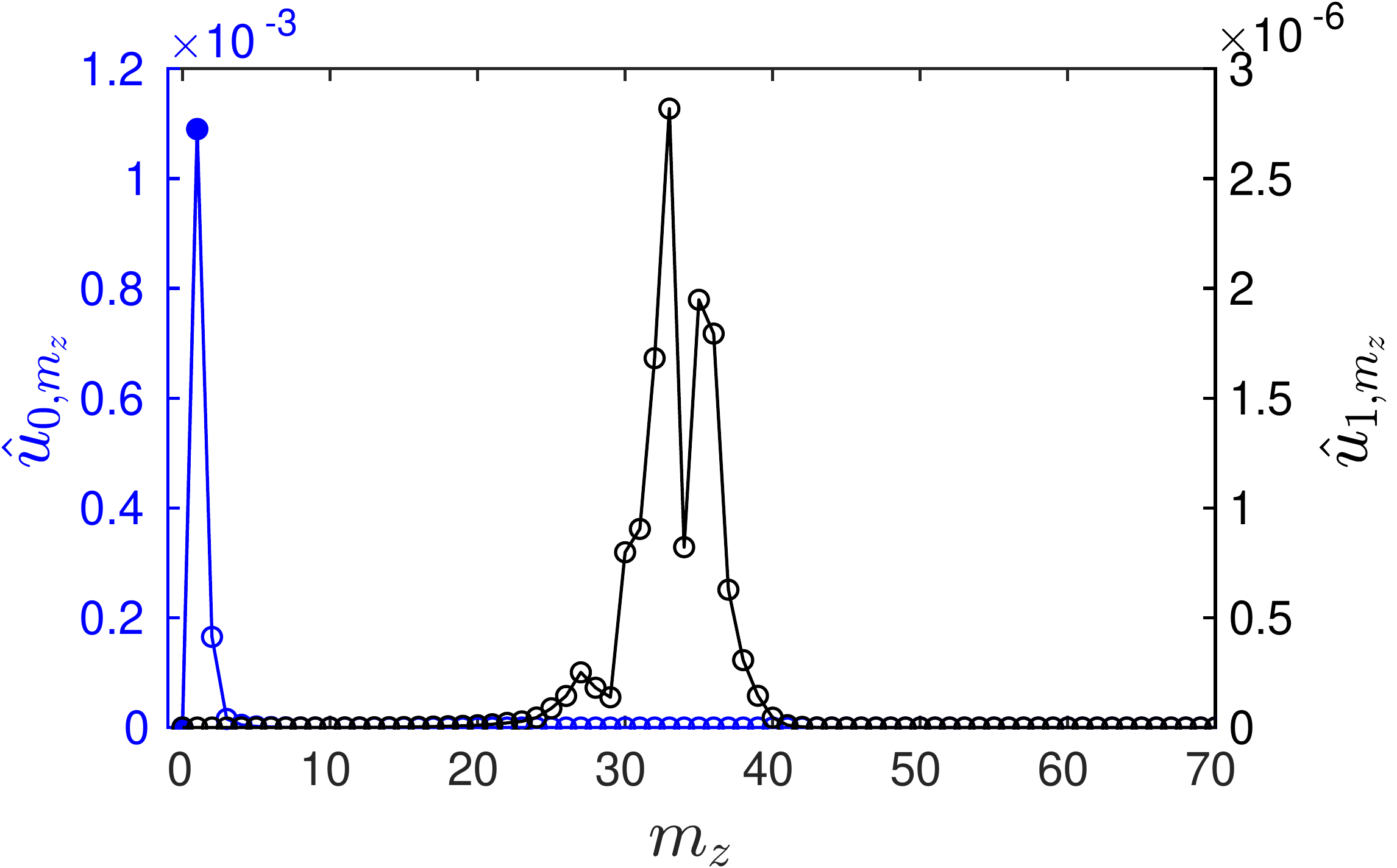}}
\caption{Illustrative Fourier spectra $\hat{u}_{0,m_z}$ and $\hat{u}_{1,m_z}$ (a) before band decay and (b) in the final relaxation to laminar flow. $Re=830$. The black symbols $\hat{u}_{1,m_z}$ with $m_z$ surrounding 35 correspond to streaks while the blue symbols $\hat{u}_{0,m_z}$ at low $m_z$ correspond to large-scale structures. Filled symbols indicate $\hat{u}_{0,1}$ and $\hat{u}_{0,2}$.
}  
\label{fig:spec_decay}
\end{figure}
\begin{figure}
        \centering
       \subfloat{\includegraphics[width=0.49\textwidth]{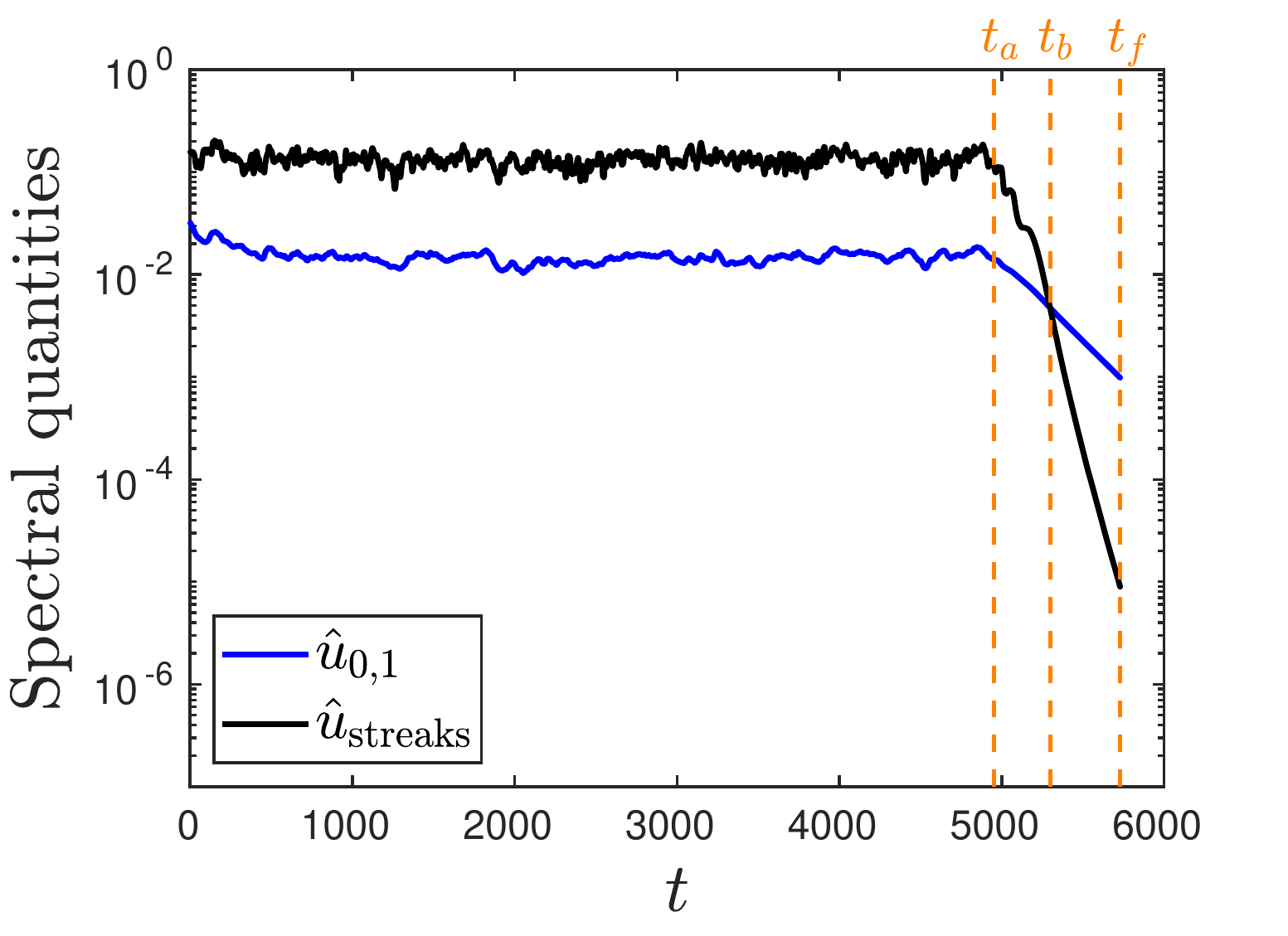}  \label{decay_specq}}
        \subfloat{\includegraphics[width=0.49\textwidth]{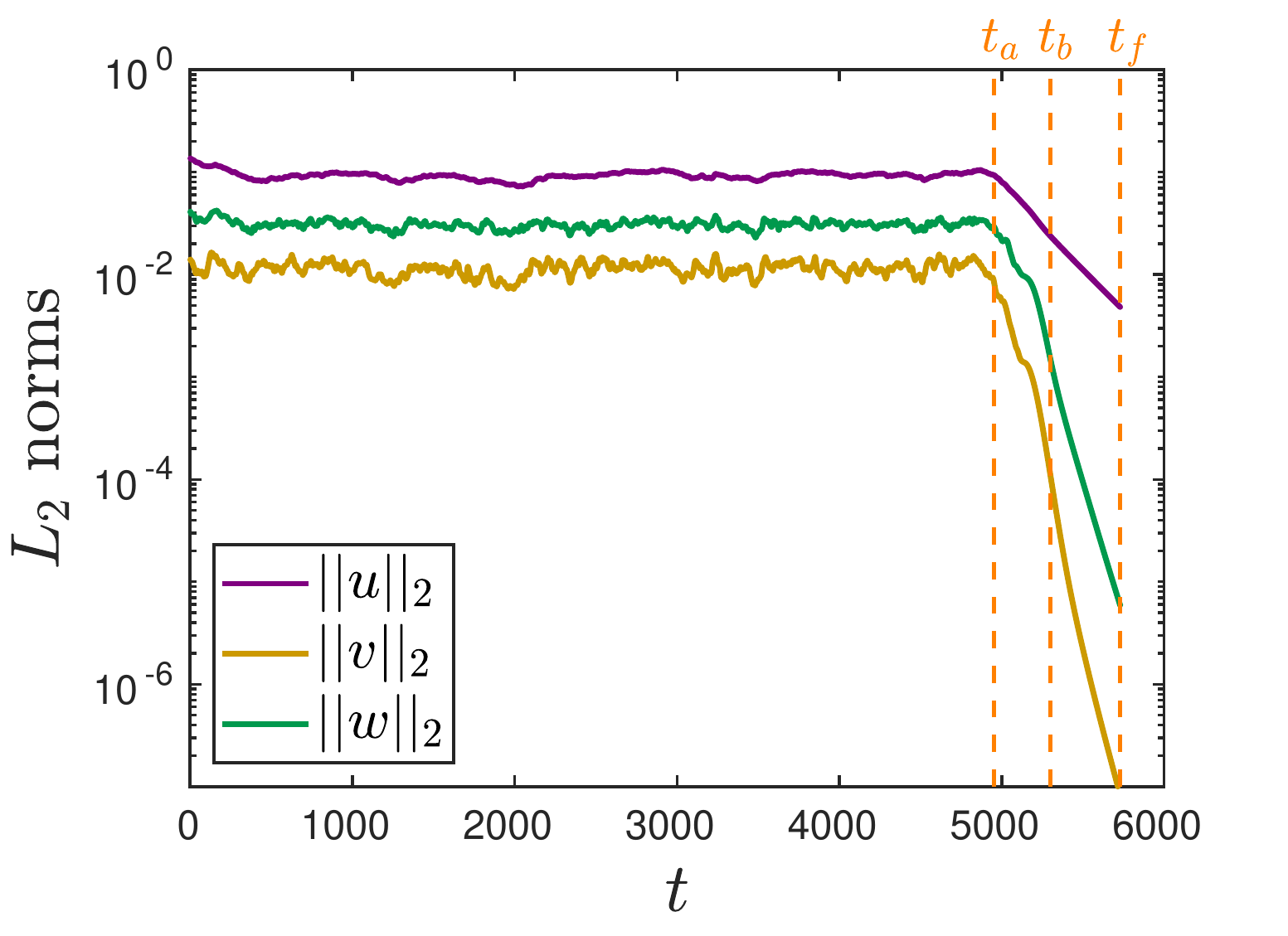}
        \label{decay_norms}}
    \caption{Time evolution of (a) spectral quantities $\hat{u}_{0,1}$ and $\hat{u}_{\text{streaks}}$, (b) $L_2$ norms $||u||_2$, $||v||_2$ and $||w||_2$ for a decay event at $Re=830$. Times $t_a$, $t_b$ and $t_f$ refer to slices shown on Fig.~\ref{fig:decay_slices}. The band starts to decay at $t_a$, $\hat{u}_{0,1} = \hat{u}_{\text{streaks}}$ at $t_b$, and the relaminarization is considered as complete at $t_f$.}
    \label{}
        \centering
       \subfloat{\includegraphics[width=0.49\textwidth]{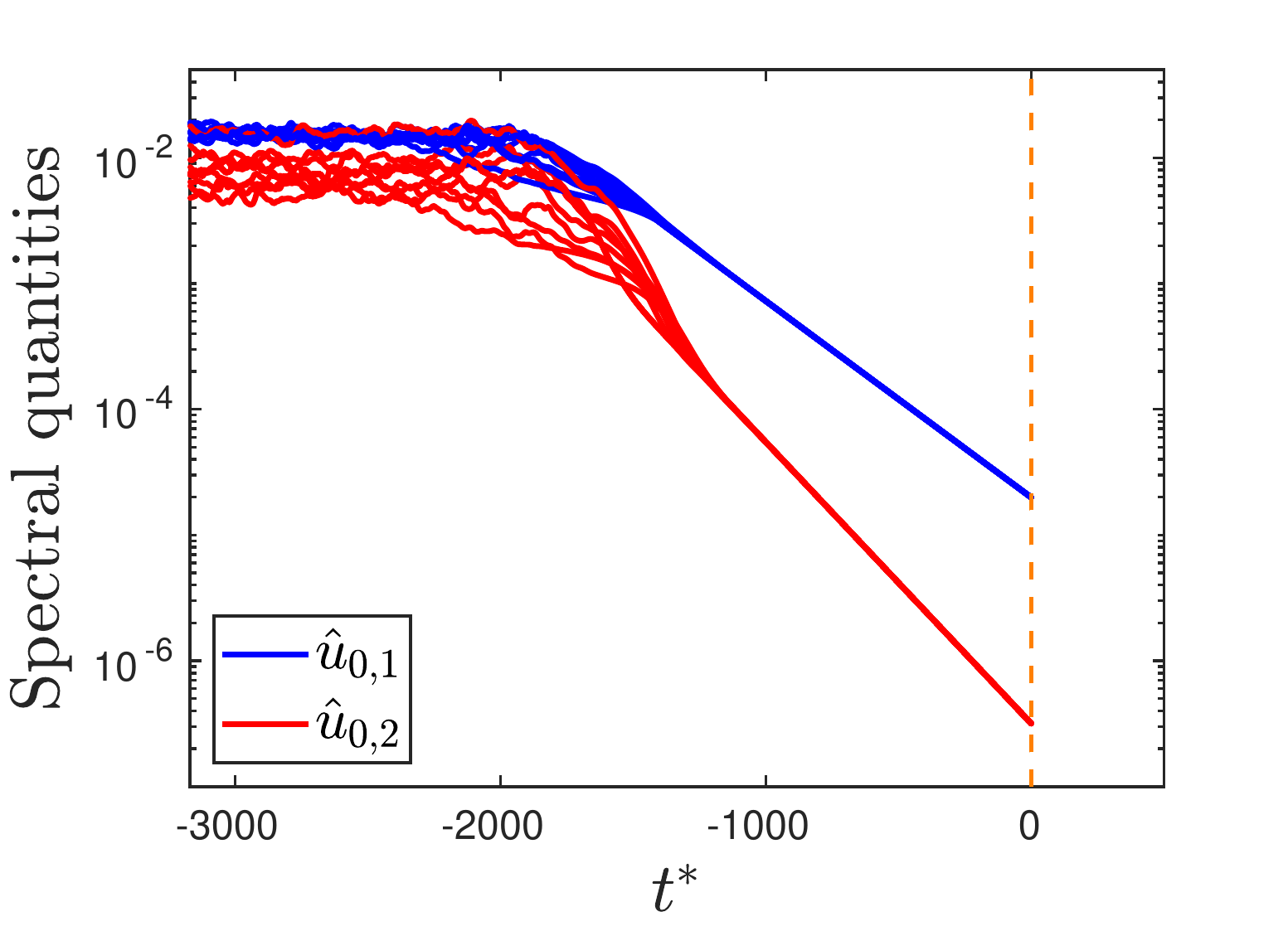}  \label{decay_specq_stoch}}
        \subfloat{\includegraphics[width=0.49\textwidth]{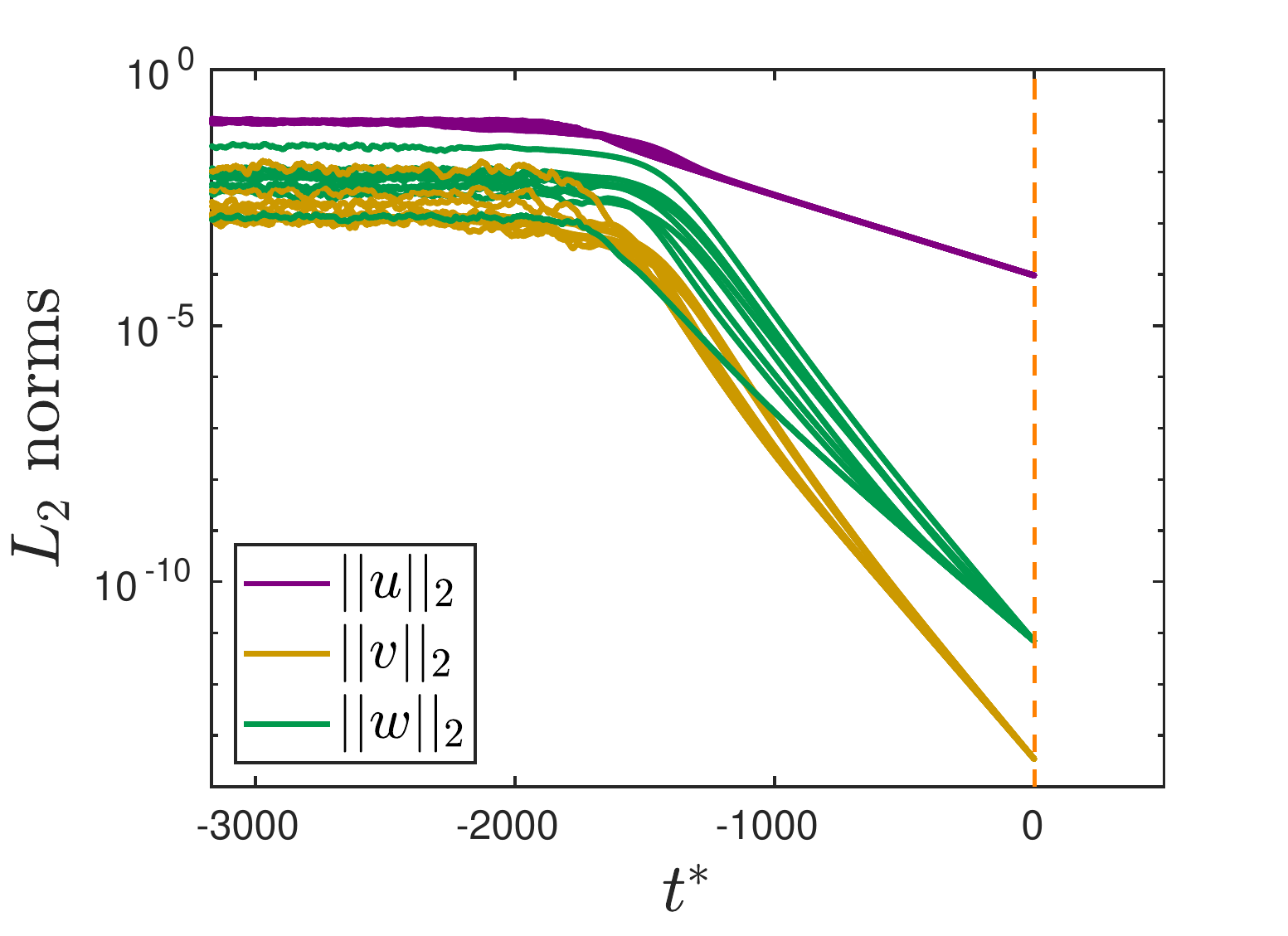}
        \label{decay_norms_stoch}}
    \caption{Time evolution of (a) $\hat{u}_{0,1}$ and  $\hat{u}_{0,2}$ and of (b) $||u||_2$, $||v||_2$ and $||w||_2$ during ten realizations of decay events at $Re=830$. Time $t^*$ and vertical quantities are respectively translated and scaled to obtain the same final value for each realization. Final decay rates for  $\hat{u}_{0,1}$ and  $\hat{u}_{0,2}$ (a) are $-3.6 \times 10^{-3}$ and $-5.2 \times 10^{-3}$, respectively.}
    \label{decay_stoch}
\end{figure}

\subsection{Decay}
\label{sec:decay}

We now focus on the decay and splitting events.
Figure \ref{fig:decay_slices}  illustrates a typical decay event, a turbulent band at $Re=830$ that persists as a long-lived metastable state before abruptly decaying to laminar flow. A visualisation of the $x$ velocity is shown in the $y=-0.8$ plane, approximately where the streaks are most intense, at representative times during the final decay to laminar flow.

States can be quantitatively characterized via their instantaneous $(x,z)$ Fourier spectra. Figure \ref{fig:2D_spec} shows an example of such a 2D Fourier spectrum of the $x$ velocity at $y=-0.8$, $Re=830$, corresponding to the snapshot $t=4850$ on Figure \ref{fig:decay_slices}. We observe that the amplitudes along horizontal lines $m_x=0$ and $m_x= \pm 1$ are much larger than the others. For brevity, we use $\hat{u}_{m_x,m_z}$ to denote the modulus of the 2D Fourier component $(\pm m_x,\mp m_z)$ of the $x$ velocity evaluated at $y=-0.8$. We recall from Eq.~\eqref{eq:specsum} that $m_x=1$ corresponds to a wavelength of $L_x=6.6$, while $m_z=1$ corresponds to a wavelength of $L_z=100$. The large-scale pattern for a single band is characterized by the $x$-constant and $z$-trigonometric Fourier coefficient $\hat{u}_{0,1}$. Streaks are the small-scale spanwise variation of the streamwise velocity. Here we use the $x$-trigonometric Fourier coefficients of the $x$-velocity as a proxy for streak amplitude:
$$
\hat{u}_{\text{streaks}} = \sum_{m_z=0}^{100} \hat{u}_{1, m_z}
$$
While the $x$ direction of the tilted domain does not correspond to the spanwise direction, it is clear from Fig.~\ref{fig:decay_slices} that the streaks correspond to $x$-wavenumber $m_x=1$.  The velocity in the $x$ direction is not the streamwise velocity, but it has a large projection in the streamwise direction.

Figure \ref{fig:spec_decay} illustrates the spectra before decay
($t_a=4950$) and near at the end of the decay process ($t_f=5700$). The final stages of the flow field as it returns to laminar flow is almost exclusively contained in the $\hat{u}_{0,1}$ coefficient corresponding to no $x$ dependence and trigonometric $z$ dependence on the scale of the simulation domain.
Weak streaks are still discernible, but their amplitudes are $10^{-3}$ that of the large-scale flow $\hat{u}_{0,1}$. (Note right-hand scale in Fig. \ref{fig:spec_decay}(b).)
This shows that the decay from a turbulent band to the laminar state results in a large-scale flow structure aligned with, and moving parallel to, the band. This large-scale flow, although weak and declining during laminarization, dominates the streak patterns characterizing turbulence.

Figure \ref{} plots the time evolution of spectral quantities and velocity norms.
The life of the band is characterized by small random fluctuations in the spectral quantities and the velocity norms, especially $\hat{u}_{\text{streaks}}$, which shows the strongest variability. After time $t=t_a=4950$, all the signals suddenly undergo exponential decay, with $||u||_2$ and $\hat{u}_{0,1}$ decaying more slowly than $||w||_2$, $||v||_2$ and $\hat{u}_{\text{streaks}}$.
Small-scale streaks and rolls have been shown to have different temporal decay rates in a Couette-Poiseuille quenching experiment \cite{liu2020}.

After the decay process begins, the averaged absolute level of the streaks $\hat{u}_{\text{streaks}}$ decays more rapidly than the large-scale component $\hat{u}_{0,1}$, resulting in the crossing of $\hat{u}_{\text{streaks}}$ and $\hat{u}_{0,1}$ at time $t=t_b=5300$ in Fig. \ref{decay_time}. From this point, the one-band structure becomes prominent in comparison with the streaks.  One sees indeed on the physical slices of Fig. \ref{fig:decay_slices} that the remaining weak flow consists primarily of an $L_z$-periodic structure, constant over $x$, and moving parallel to the previous band. Band-orthogonal and cross-channel velocities $w$ and $v$ are negligible in comparison to $u$, and only show a remaining streaky pattern.


We now consider how these quantities vary for different decay events. 
Figure \ref{decay_stoch} presents the evolution of spectral quantities and velocity field norms for 10  decay events. For each realization $i$, time is translated, $t^* = t-t_{f,i}$, so that all realizations end at the same time: $t^* =0$. 
Quantities are also normalized to obtain the same final value:  $q^* = \text{min}(q_{f,i})\times q_i/q_{f,i}$. Note that the final time for the simulation $t_{f}$ is dictated by the criterion $||u||_2<5\times 10^{-3}$ and that $||u||_2$ is dominated by $\hat{u}_{0,1}$, which is why both signals terminate with the same final value for each realization. 

The evolution of the spectral component $\hat{u}_{0,1}(t)$ 
for the different realizations all eventually collapse onto a single curve. The same is true, slightly later, for $\hat{u}_{0,2}(t)$.
These final phases of the evolution correspond to viscous diffusion; $\hat{u}_{0,1}(t)$ and $\hat{u}_{0,2}(t)$ evolve towards eigenvectors of laminar plane channel flow.
The difference between their decay rates (eigenvalues) is due to differences in their cross-channel dependence.

The norm $||u||_2$ also behaves in this way, since it is dominated by $\hat{u}_{0,1}$, but $||v||_2$ and $||w||_2$ do not. These are sums over different spectral components each with its own decay rate, and the levels of these components differ from one realization to the next, thereby leading to different decay rates for each realization.

\subsection{Splitting}

\begin{figure}
        \centering
        \subfloat{\includegraphics[width=0.8\textwidth]{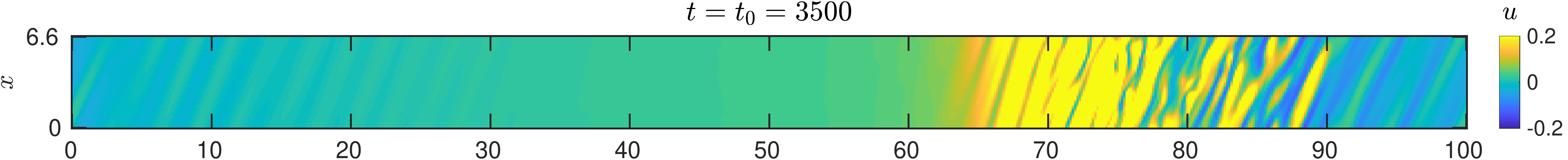}}\\
        \vspace*{-0.78em}
        \subfloat{\includegraphics[width=0.8\textwidth]{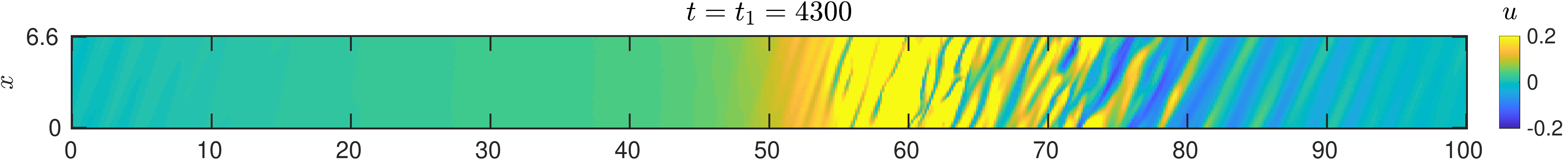}}\\       
        \vspace*{-0.78em}
        \subfloat{\includegraphics[width=0.8\textwidth]{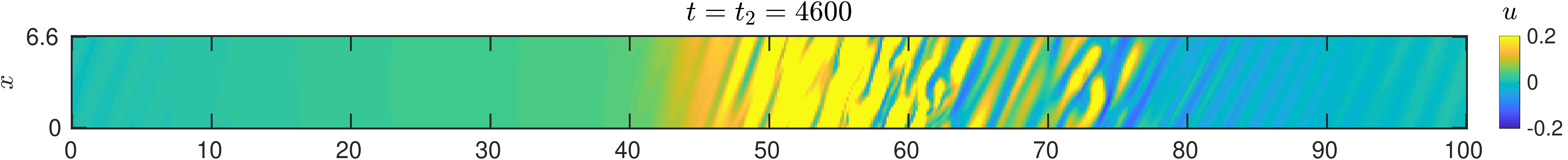}}\\        
        \vspace*{-0.78em}
        \subfloat{\includegraphics[width=0.8\textwidth]{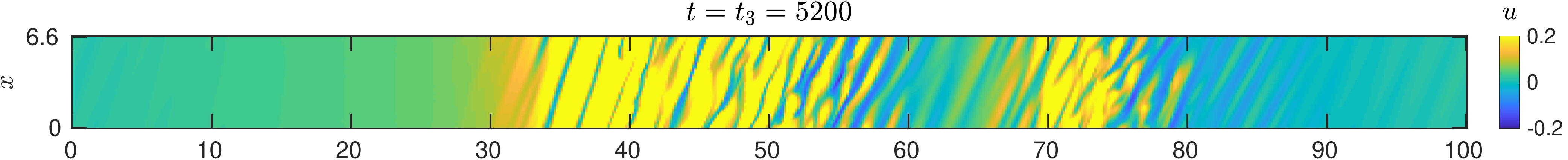}}\\
        \vspace*{-0.78em}
        \subfloat{\includegraphics[width=0.8\textwidth]{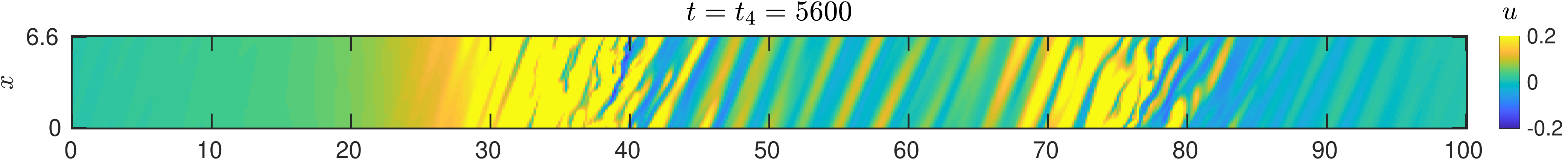}}\\
        \vspace*{-0.78em}
        \subfloat{\includegraphics[width=0.8\textwidth]{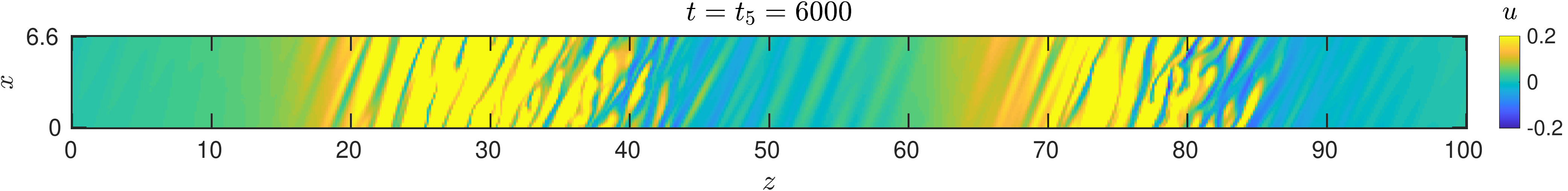}}\\
    \caption{Band splitting at $Re=1200$. Plotted is the $x$ velocity in $(x,y)$ planes at $y=-0.8$.}
    \label{fig:split_xz}
\end{figure}

\begin{figure}
        \centering
        \subfloat[]{\includegraphics[width=0.2\textwidth]{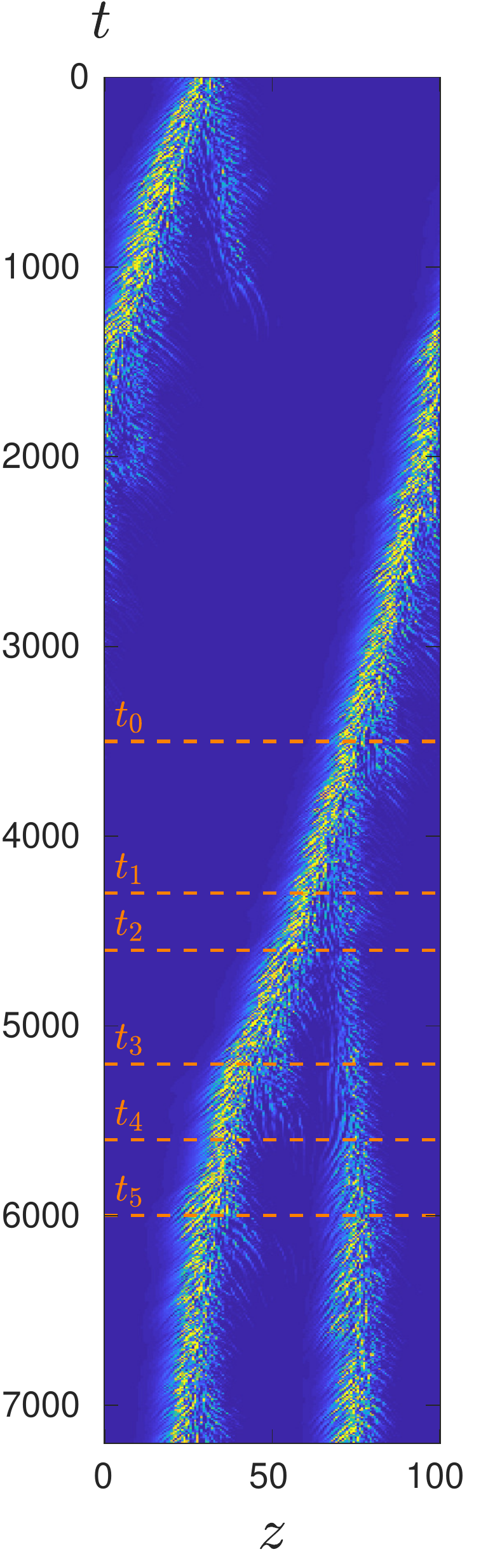} \label{split_probes}}
        \subfloat[]{\includegraphics[width=0.2\textwidth]{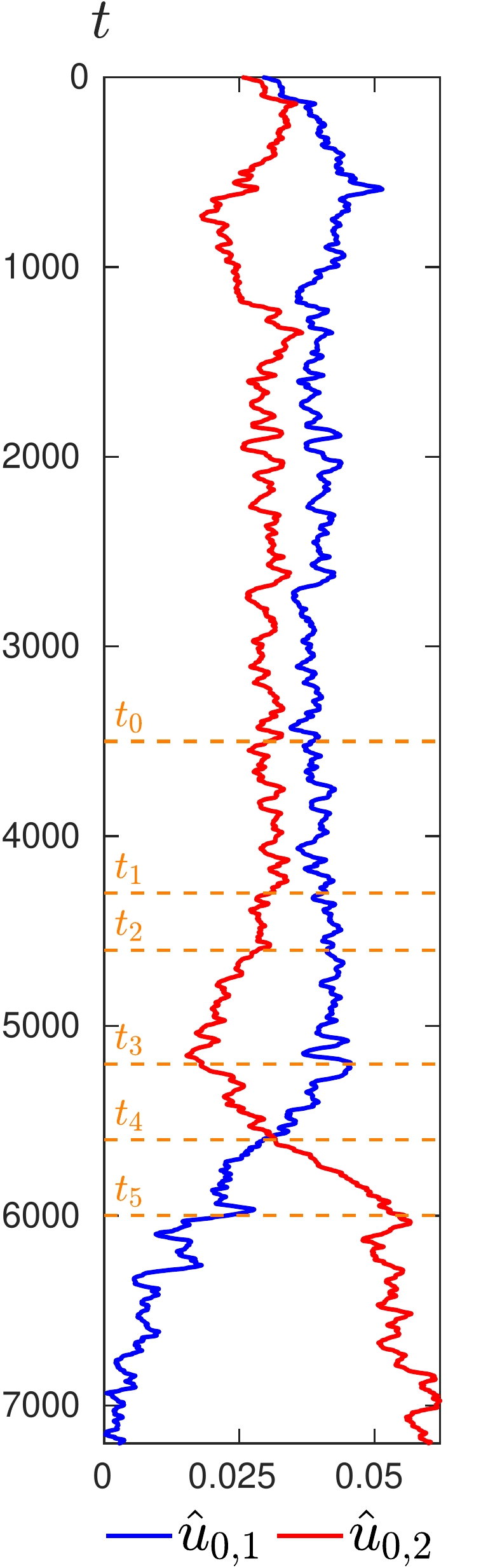} \label{spec_q}}
        \subfloat[]{\includegraphics[width=0.2\textwidth]{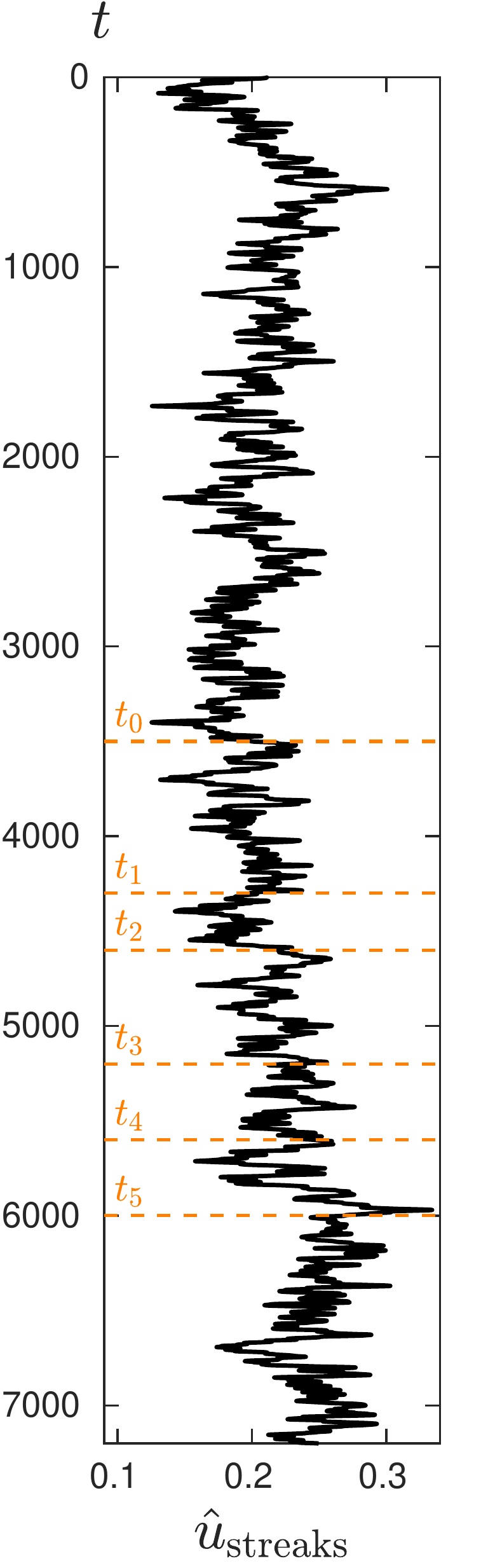} \label{ustreaks}}
        \subfloat[]{\includegraphics[width=0.2\textwidth]{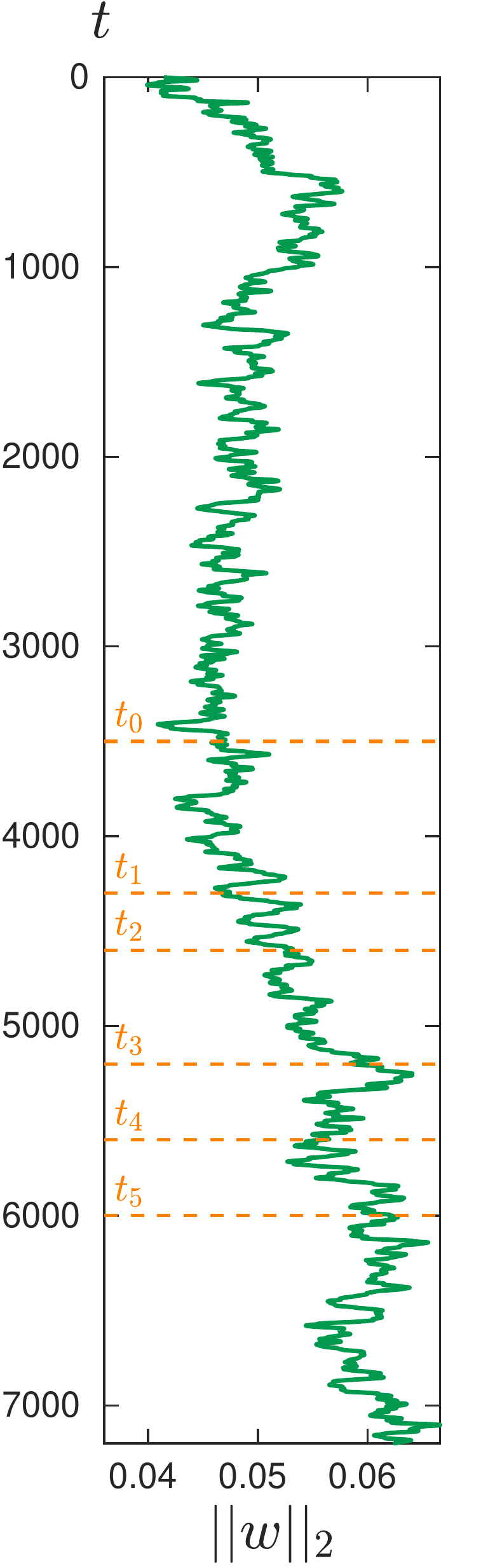} \label{w2}}
    \caption{Evolution of a band while it splits at $Re=1200$. (a) Spatiotemporal diagram of the band. Colors show the turbulent perturbation energy $E$ between 0 (blue) and 0.1 (yellow). (b, c, d) Time evolution of spectral quantities $\hat{u}_{0,1}$ and $\hat{u}_{0,2}$ (b), $\hat{u}_{\text{streaks}}$ (c) and the $L_2$-norm $||w||_2$ (d).}
    \label{fig:split_time}
\end{figure}
\begin{figure}
        \centering
    \subfloat[$t_0=3500$]{\includegraphics[width=0.5\textwidth]{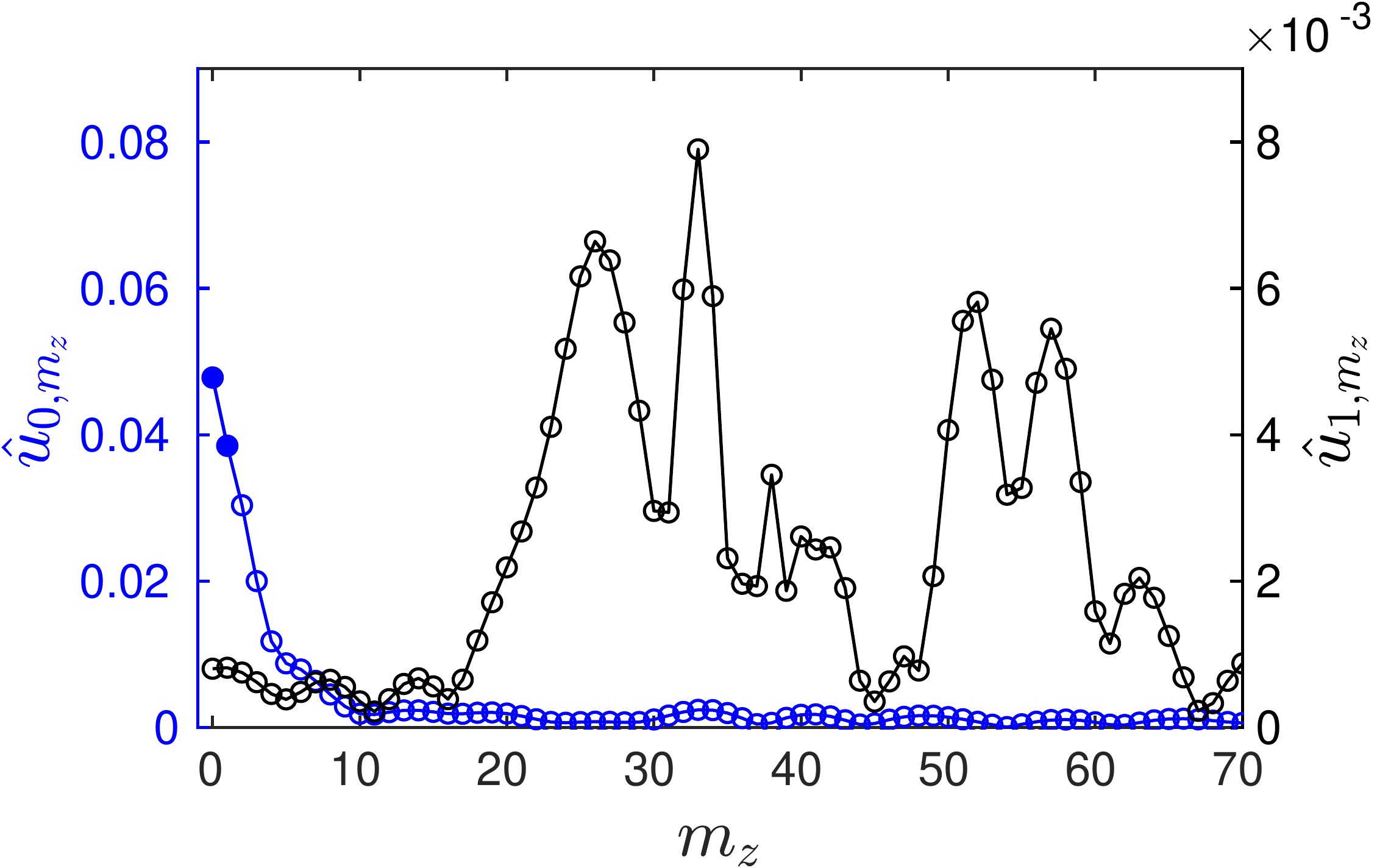}}
        \subfloat[$t_5=6000$]{\includegraphics[width=0.5\textwidth]{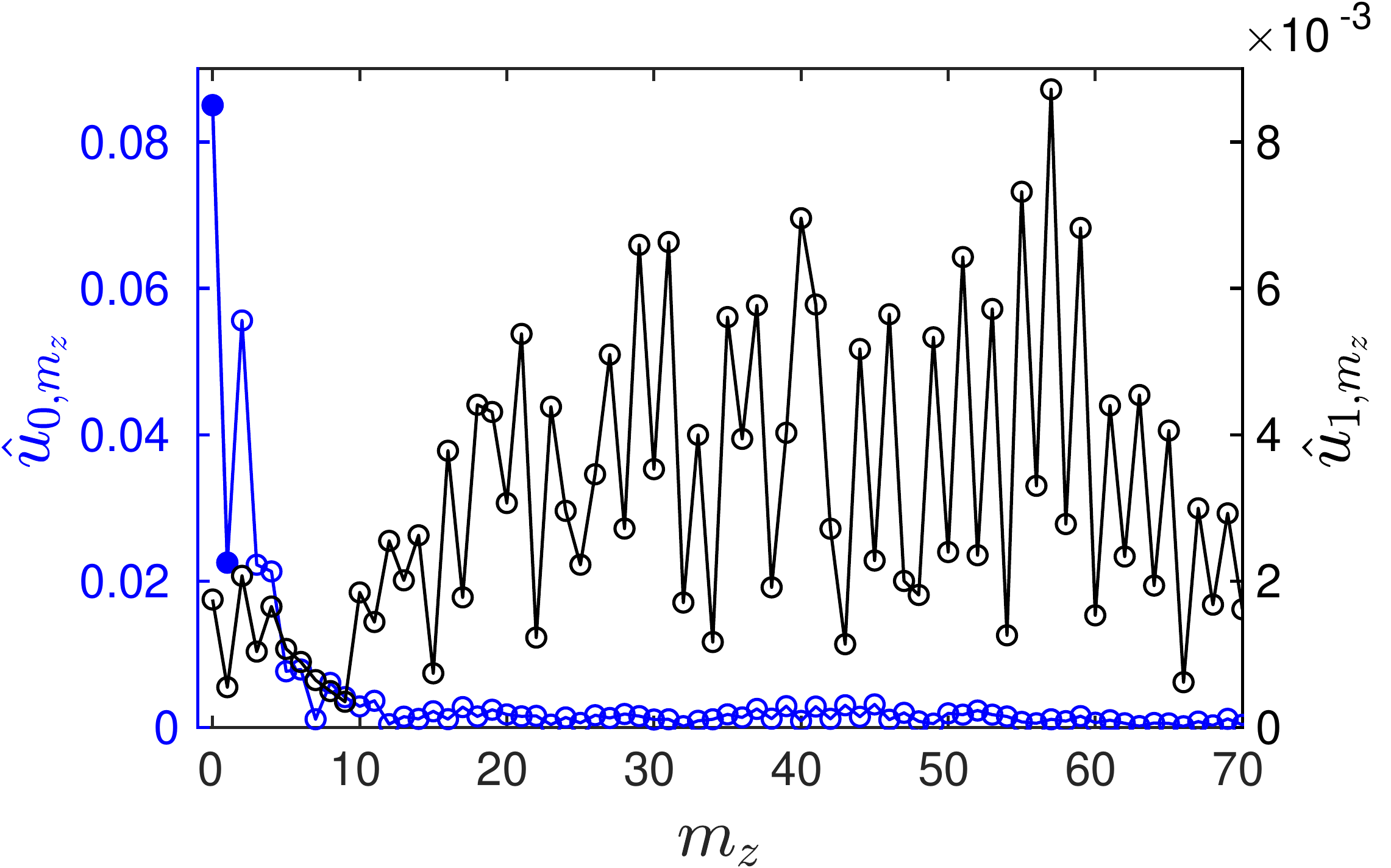}}
    \caption{Illustrative Fourier spectra $\hat{u}_{0,m_z}$ and $\hat{u}_{1,m_z}$ (a) before and (b) after band splitting at $Re=1200$. The black symbols $\hat{u}_{1,m_z}$ with $m_z$ surrounding 35 correspond to streaks while the blue symbols $\hat{u}_{0,m_z}$ at low $m_z$ correspond to large-scale structures. Filled symbols indicate $\hat{u}_{0,1}$ and $\hat{u}_{0,2}$.}
    \label{fig:spec_split}
\end{figure}

A splitting event at $Re=1200$ is shown in Fig. \ref{fig:split_xz} via the evolution of $(x,z)$ slices of $u$, at  times from $t_0$ (initial band) to $t_5$. The turbulent band at $t_1=4300$ is wider than it is at $t_0=3500$.  At $t_2=4600$ one sees the appearance of a gap in the turbulent region corresponding to the birth of the second band. 
The parent band continues to move towards lower $z$ while the child band remains at its position and intensifies from $t_2$ to $t_5$, smoothly acquiring all the characteristics of the parent band.


Figure \ref{fig:split_time} presents a spatio($z$)-temporal diagram of the perturbation energy and traces the evolution of spectral quantities $\hat{u}_{0,1}$ and  $\hat{u}_{0,2}$ at $y=-0.8$, which represent a single or a double banded pattern. The evolution of $\hat{u}_{\text{streaks}}$ and of the $L_2$-norm $||w||_2$ are also shown.
A slight initial drop in the two-band coefficient $\hat{u}_{0,2}$ is seen from $t= t_1 = 4300$, which coincides with the appearance of the second band. A laminar gap opens between the initial band and its offspring at $t=t_2=4600$. Then $\hat{u}_{0,2}$ starts to increase whereas $\hat{u}_{0,1}$ decreases, from $t=t_3= 5200$. The two quantities cross at $t=t_4= 5600$ and finally reach plateaus at $t=t_5=6000$. This is the time from which the energy of the second band reaches approximately the same level as that of the first band, as seen from the spatio-temporal diagram (Fig. \ref{split_probes}).
The other quantities, $\hat{u}_{\text{streaks}}$ and $||w||_2$, follow slightly different trends from those of the spectral coefficients, as shown on Fig. \ref{ustreaks} and \ref{w2}. Oscillations in $\hat{u}_{\text{streaks}}$ are strong and it is difficult to distinguish trends corresponding to the band evolution. However, there is a relatively strong increase in the streak intensity just before $t_5$, when the second band is fully developed. In addition, $||w||_2$ increases from $t_1$ to $t_3$ and then reaches a plateau of around 0.06.

The evolution before the splitting shows a missed splitting event between $t=200$ and 1000. A weakly turbulent patch detaches from the initial stripe, and quantities $\hat{u}_{0,1}$, $\hat{u}_{0,2}$, $\hat{u}_{\text{streaks}}$, and $||w||_2$ all follow a trend between $t=200$ and 600 similar to that  between $t_2$ and $t_3$.  The birth ceases after $t=1000$: $\hat{u}_{0,2}$ does not increase sufficiently to cross $\hat{u}_{0,1}$, and $\hat{u}_{\text{streaks}}$ and $||w||_2$ drop to their previous levels.

Figure \ref{fig:spec_split} shows a comparison between Fourier spectra $\hat{u}_{0,m_z}$ and $\hat{u}_{1,m_z}$ before and after splitting. 
The decrease in $\hat{u}_{0,1}$ and increase in $\hat{u}_{0,2}$, already seen in Fig.~\ref{fig:split_time}b, appears clearly. 
In addition, the two-band streak spectrum $\hat{u}_{1,m}$ shows conspicuous small-scale oscillations due to the fact that a perfectly $L_z/2$-periodic field would contain only even modes.

We now carry out simulations, still at $Re=1200$, in a shorter tilted domain of length $L_z=50$ to avoid secondary splittings which would lead to a three-band state. All realizations of the formation of the second band follow the same sequence of events previously described. 
Meanwhile, the three-band component $\hat{u}_{0,3}$ can also be monitored to analyze the interactions between modes 1 and 2 during the splitting. 

\begin{figure}
        \centering
       \includegraphics[width=0.75\textwidth]{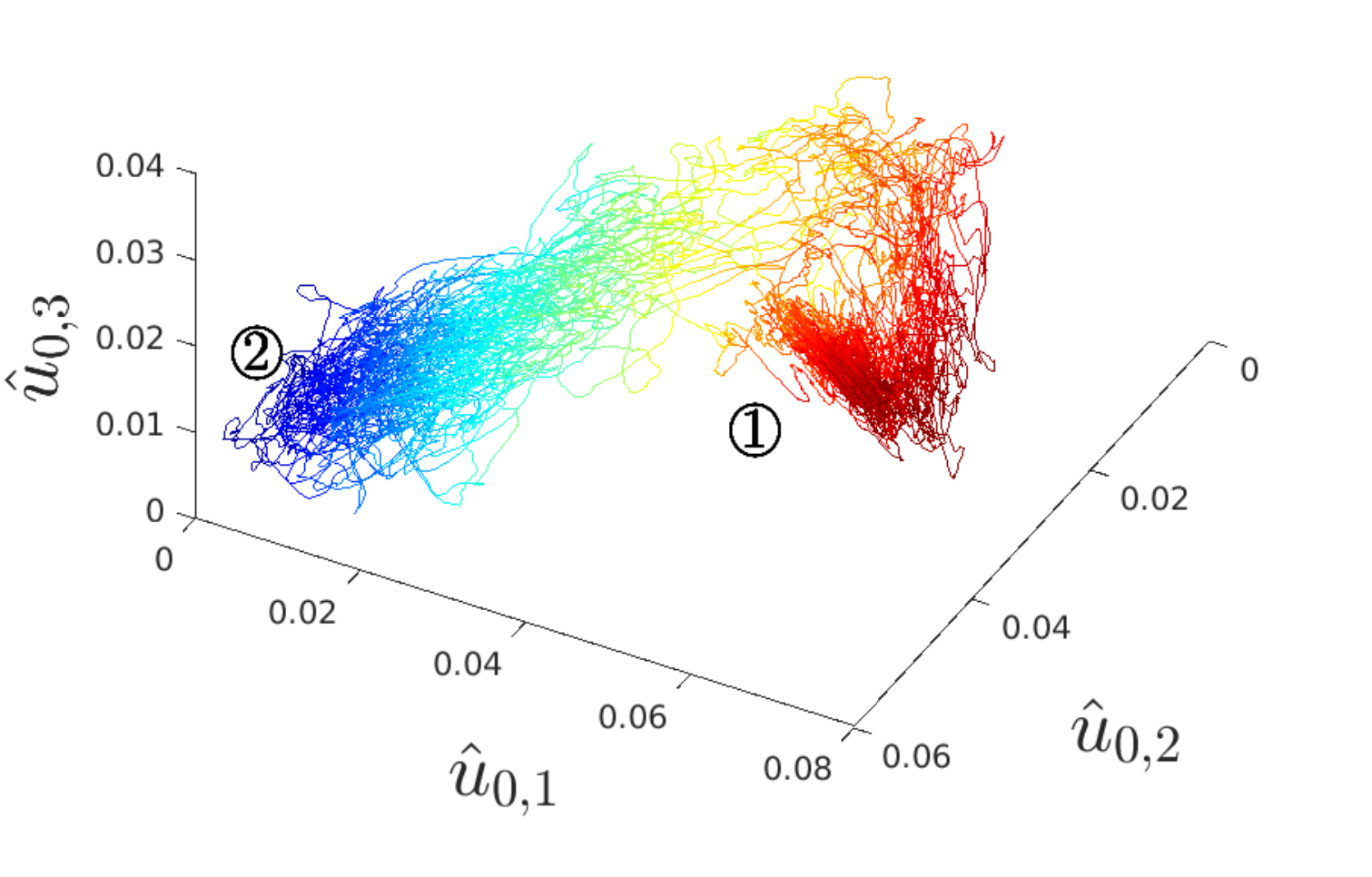}
    \caption{Evolution of spectral quantities during 10 splittings at $Re=1200$, in a domain of length $L_z=50$. Each curve represents one simulation, and is colored by $\hat{u}_{0,1}$ to illustrate the transition between a one-band (1) to a two-band state (2).}
    \label{fig:split_path}
\end{figure}

This evolution is represented in a phase portrait $(\hat{u}_{0,1},\hat{u}_{0,2},\hat{u}_{0,3})$ in Fig. \ref{fig:split_path}. The one-band state is characterized here by an average segment around which the spectral components show noisy oscillations (state 1) because of the proportionality between the components.
Because the two-band state selects the even components (see Fig.~\ref{fig:spec_split}b), $\hat{u}_{0,1}$ and $\hat{u}_{0,3}$ have low values and show no correlation with the prominent $\hat{u}_{0,2}$. 
This representation shows that large-scale spectral components statistically follow the same transition path from one to two turbulent bands. This common transition path can be seen as a low-dimensional projection of the dynamics of band splitting. 
Such a statistical pathway for configuration changes in a turbulent fluid system was observed in the case of barotropic jet nucleation \cite{bouchet2019rare}.

\section{Statistics of band decay and splitting}
\label{sec:stochastic}

\begin{figure}
        \centering
       \subfloat{\includegraphics[width=0.49\textwidth]{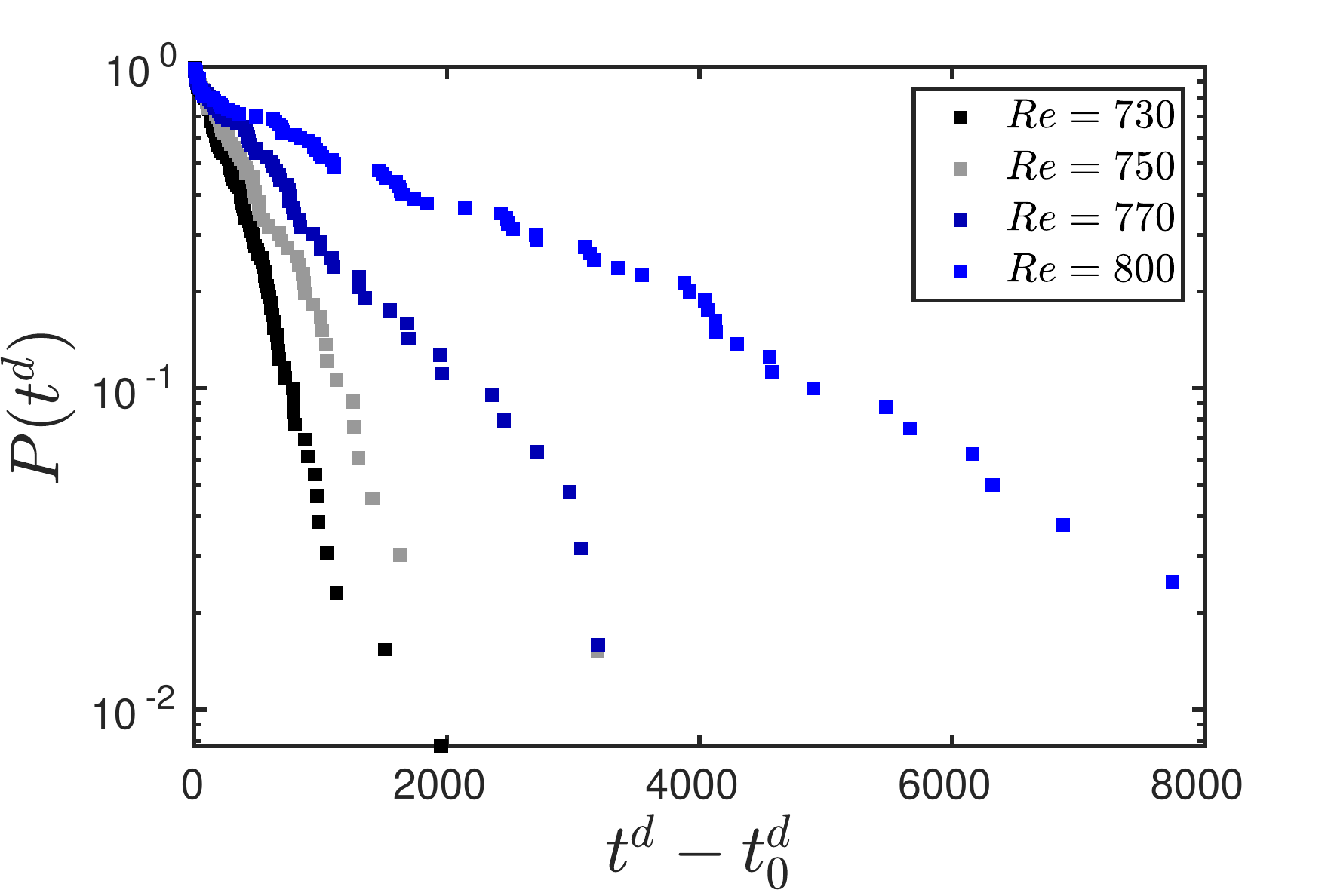}  \label{Stat_decay1}}
        \hspace*{-0.5em}
        \subfloat{\includegraphics[width=0.49\textwidth]{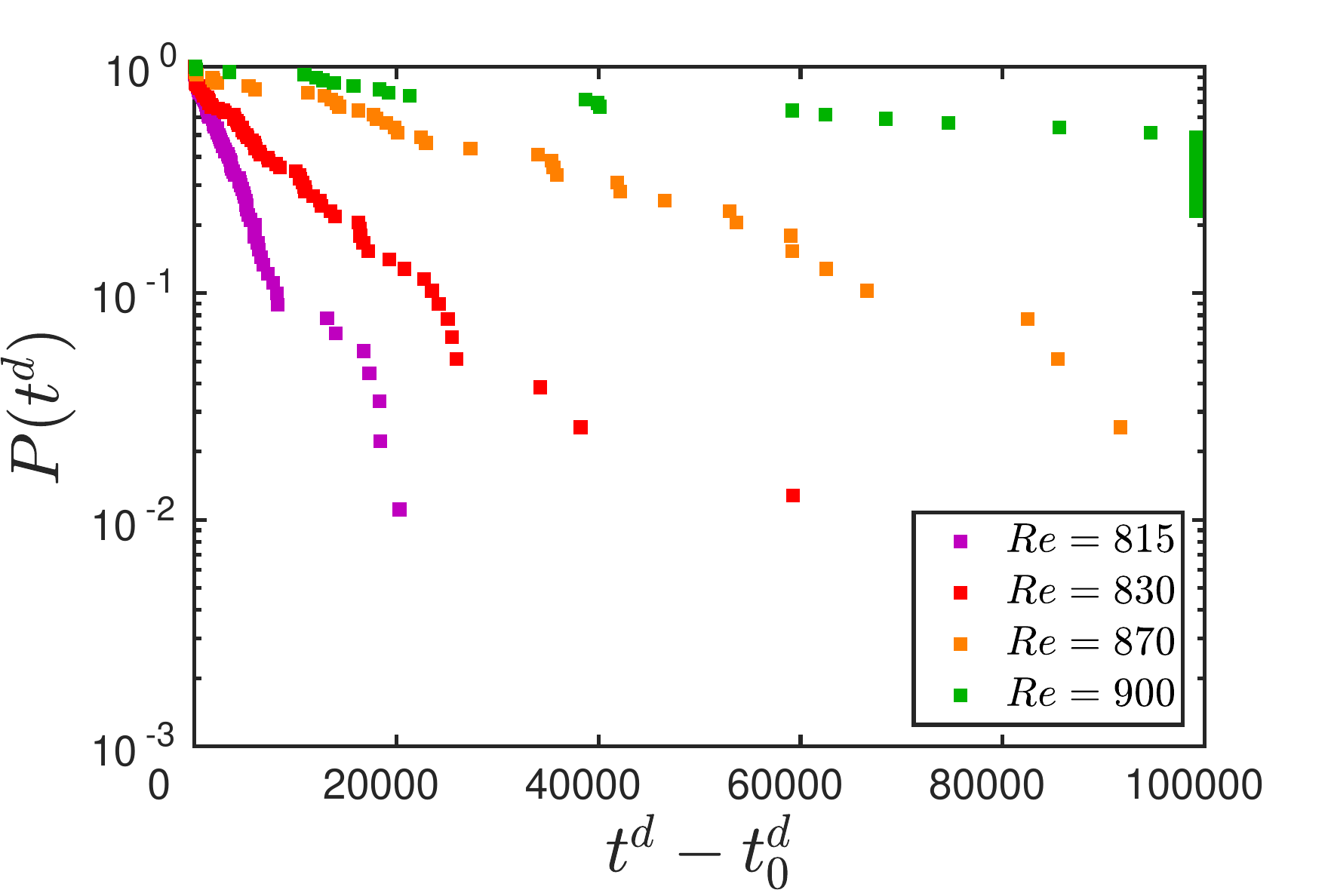}
        \label{Stat_decay2}}
    \caption{Survival probability distributions for the decay of a turbulent band, $Re \in [730, 900]$.}
    \label{fig:stat_decay}
     \subfloat{\includegraphics[width=0.49\textwidth]{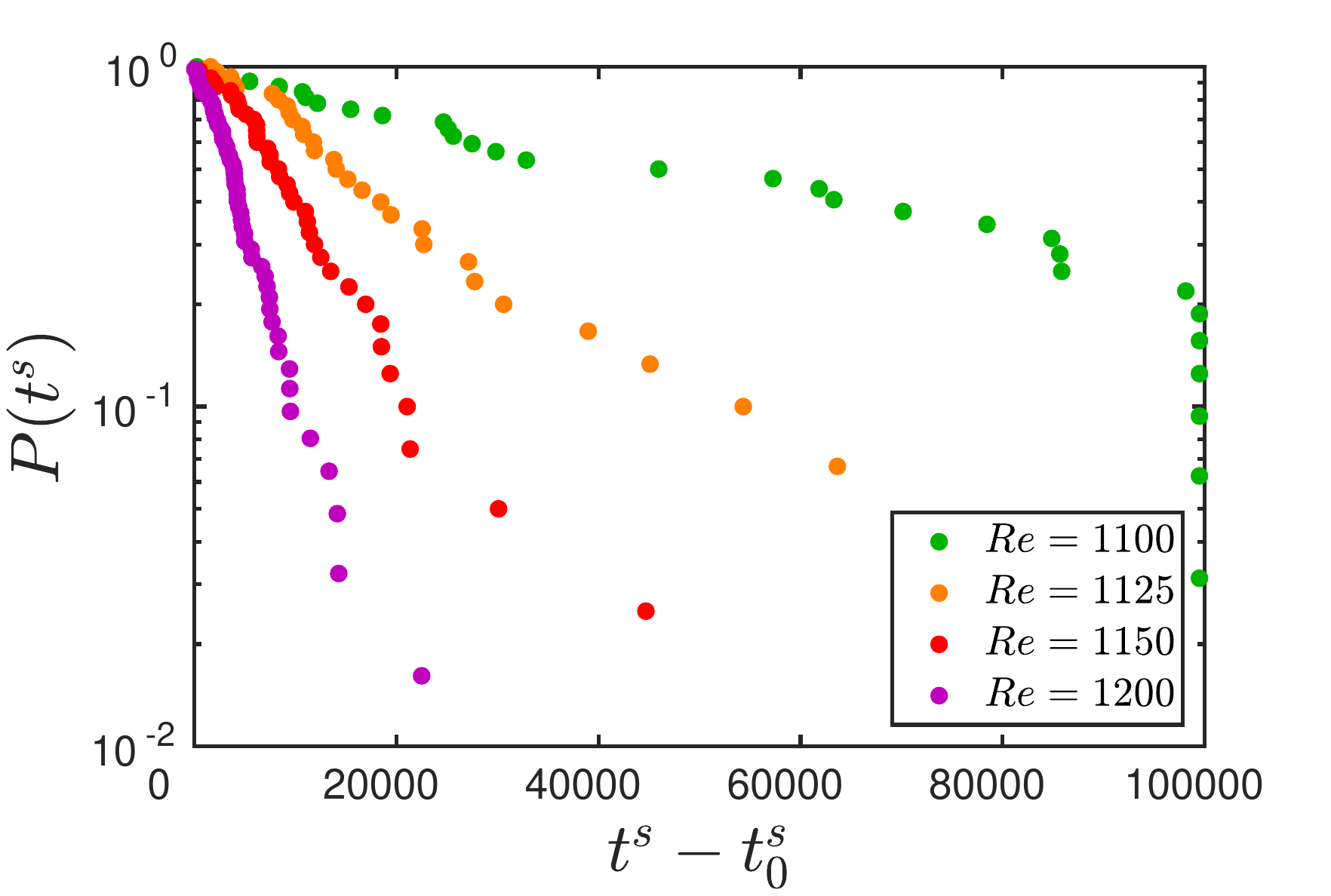}  \label{Stat_split1}}
      \hspace*{-0.5em}
        \subfloat{\includegraphics[width=0.49\textwidth]{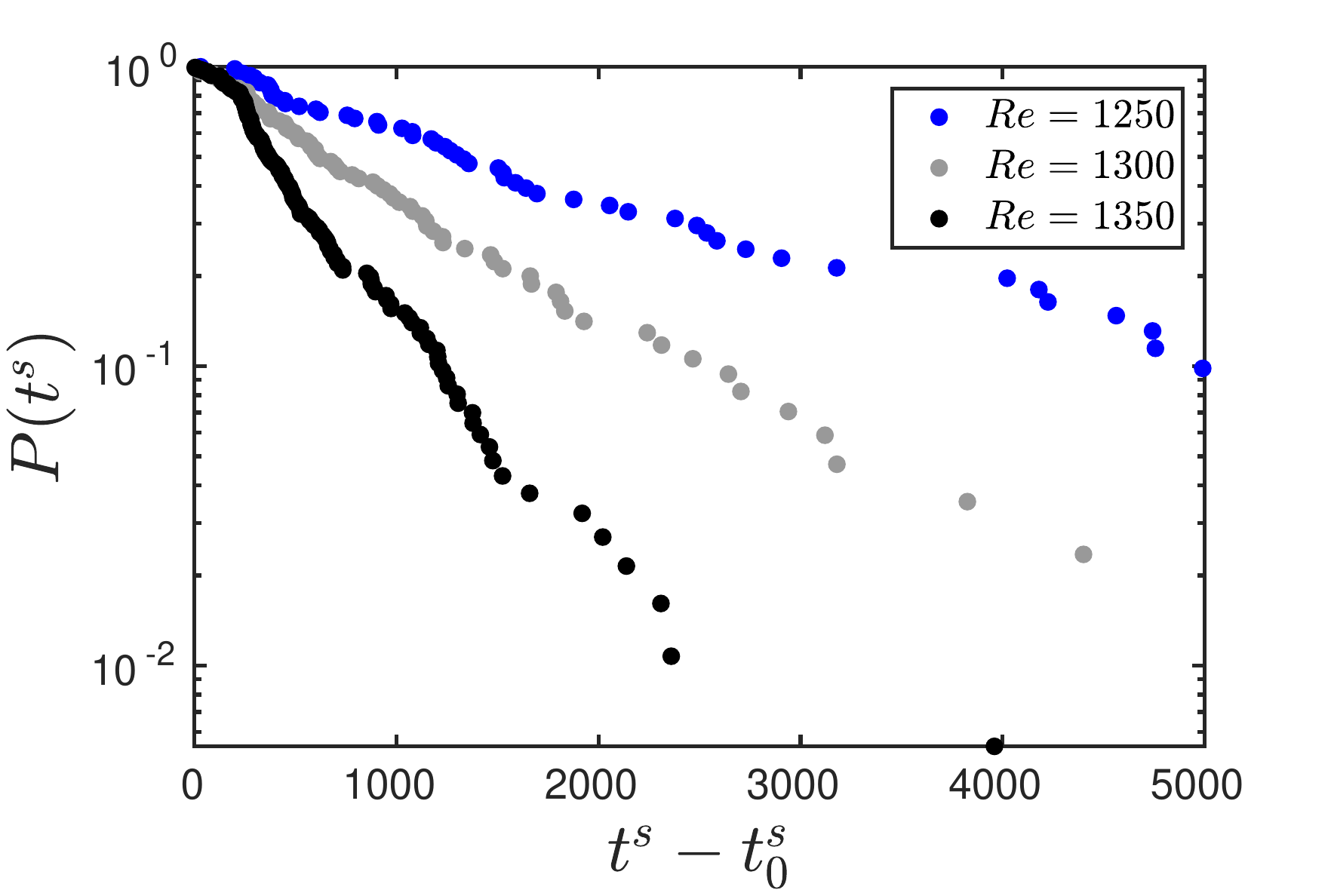}
        \label{Stat_split2}}
    \caption{Survival probability distributions for the splitting of a turbulent band, $Re \in [1100, 1350]$.}
    \label{fig:stat_split}
\end{figure}
\begin{figure}
    \centering
    \includegraphics[width=0.6\textwidth]{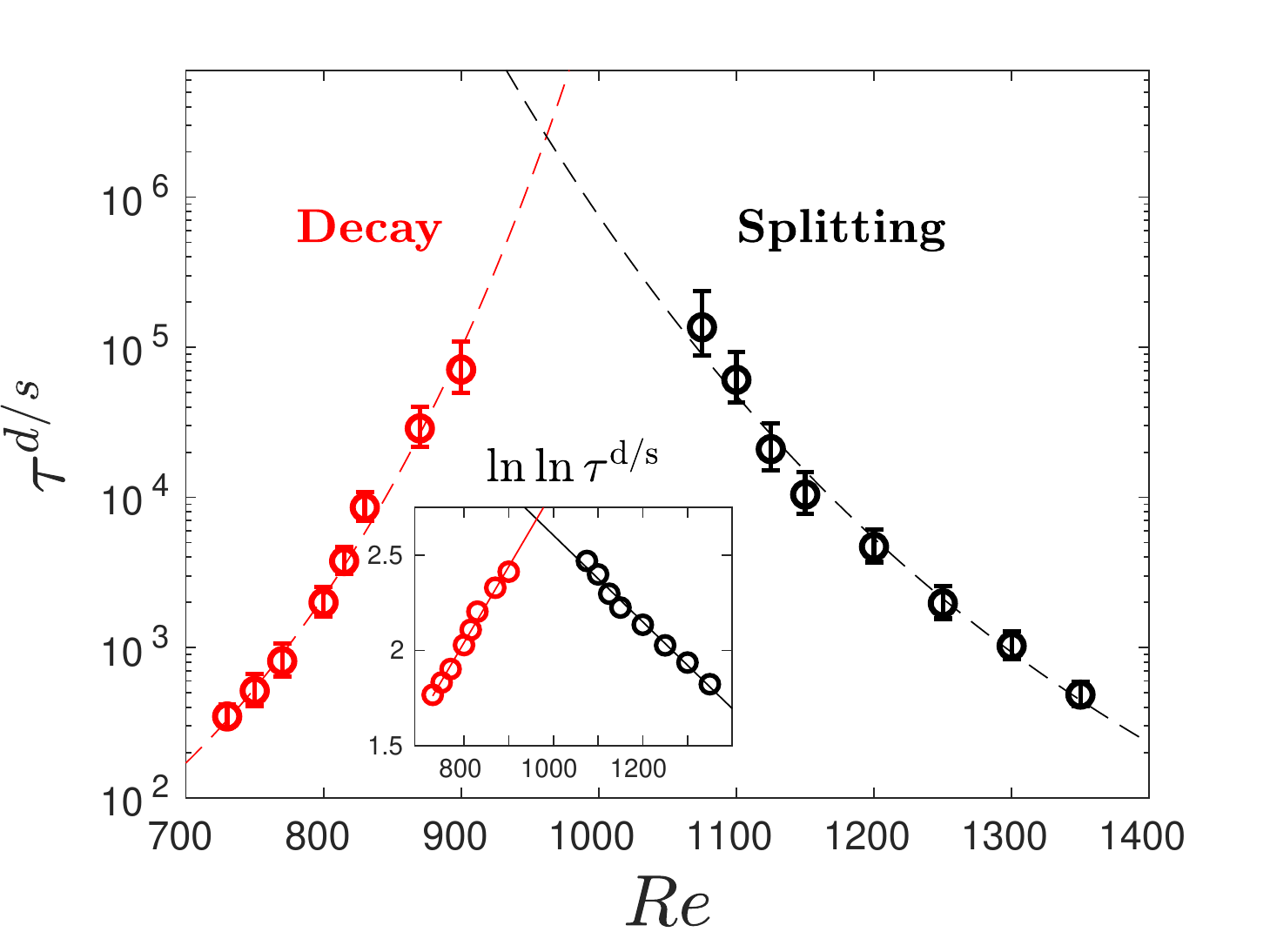}
    \caption{Variation of mean decay times (red) and splitting times (black) with Reynolds number $Re$. The error bars correspond to $95\%$ confidence intervals. Inset: $\ln \ln{\tau^{s/d}}$ versus $Re$ and associated linear fits. The crossing point is at $Re_{\rm cross}\approx 965$, $\tau\approx 3\times 10^6$.}
    \label{fig:stat_threshold}
\end{figure}

We now investigate the decay and splitting statistics of single turbulent bands over a range of Reynolds numbers. 
The mean lifetime of decay increases with $Re$, that of splitting decreases with $Re$, and hence these lifetimes are equal at some Reynolds number. The primary goal here is to determine at which Reynolds number value this occurs.
The domain size is fixed at $L_z=100$. 
Since decay and splitting events are effectively statistical, many realisations are necessary to determine the mean decay and splitting times. 
Regarding the evolution of band interactions with $L_z$ (Section \ref{sec:band_velocity}), $L_z=100$ was chosen as a compromise between mitigating the potential effect of interactions on decay and splitting probabilities and the numerical cost of a statistical study. The effect of inter-band distance on mean decay and especially on splitting times still remains an open question.
To generate large numbers of initial conditions for these realisations, we start from featureless turbulent flow at $Re=1500$ and reduce $Re$ to an intermediate value in $[900, \, 1050]$, where a single band then forms. We continue these simulations and extract snapshots, that are then used as initial conditions for simulations with $Re\in [700, \, 1350]$. 

Each simulation is run with a predefined maximum cut-off time $t_{\rm f}=10^5$. If a decay or splitting event occurs before $t_{\rm f}$, the run is automatically terminated after the event and the time is recorded. For a decay, the termination criterion is $||\mathbf{u}||_{L_2} < 0.005$, meaning that the flow has nearly reached the laminar base flow. For splitting, termination occurs when two (or more) well-defined turbulent zones 
(whose $x$ and short-time averaged turbulent energy exceed 0.005) 
coexist over more than $2000$ time units. We can then estimate the real time at which the splitting event occurs, defined as the time at which a second laminar gap appears from the initial band, through careful observations of space-time diagrams. 

For a given value of $Re$, let $N^d$, $N^s$, and $N$ be the number of decay events, splitting events, and the total number of runs, respectively. Thus $N-N^d - N^s$ is the number of runs reaching the cut-off time $t_{\rm f}$ without having decayed or split.

We consider first the decay statistics. (The splitting statistics follow similarly.) The analysis closely follows previous work; see especially \cite{avila2010transient,avila2011onset,shi}. The decay times at a given $Re$ are sorted in increasing order, giving the sequence $\{t^d_i\}_{1\leq i \leq N^d}$. The survival probability that a band has not decayed by time $t^d_i$ is then approximated by:
\begin{equation}
    P(t^d_i) = P(\text{decay at } t \geq t^d_i) =  1 - (i-1)/N.
\end{equation}

The survival distributions for decay events over a range of $Re$ are plotted on semi-log axes in Fig.~\ref{fig:stat_decay}. The data support exponential form $P(t^d_i)=\exp(-(t^d_i - t^d_0)/\tau^d(Re))$, where $\tau^d(Re)$ is the Reynolds-number-dependent mean lifetime (characteristic time) for decay and $t_0^d$ is an offset time, for $Re\geq 750$. (The case $Re=730$ exhibits deviations from an exponential distribution very similar to those observed in pipe flow at $Re=1700$ \cite{avila2010transient}). These exponential survival distributions are indicative of an effectively memoryless process, as has been frequently observed for turbulent decay in transitional flows \cite{darbyshire1995transition,faisst2004sensitive,hof2006finite,peixinho2006decay,willis2007critical, avila2010transient}.

Quantitatively, the characteristic time $\tau^d(Re)$ is obtained by the following Maximum Likelihood Estimator \cite{avila2010transient}:
\begin{equation}
    \tau^d \simeq  \frac{1}{N^{'d}}  \Big ( \sum_{t_i^d>t_0^d} (t^d_i-t^d_0)  +  (N - N^d) (t_{\rm f}-t^d_0)\Big )
    \label{eq:tau_d}
\end{equation}
where $N^{'d}$ is the number of decay events taking place after  $t_0^d$. The offset time $t_0^d$ is included to account for the time necessary for the flow to equilibrate following a change in $Re$ associated with the initial condition, and also the fixed time it takes for the flow to achieve the termination condition after it commences decay (as seen in Fig.~\ref{decay_norms}). As in \cite{avila2010transient}, we determine the value of $t_0^d$ by varying it in Eq.~\eqref{eq:tau_d}, monitoring the resulting characteristic time $\tau^{d}$, and choosing $t_0^d$ to be the minimal time for which the estimate $\tau^{d}$ no longer depends significantly on $t_0^d$. We find $t_0^d = 850$ is a good value over the range of $Re$ investigated.

The same procedure has been applied to the splitting events. The splitting times are denoted $\{t^s_i\}_{1\leq i \leq N^s}$, the estimated mean lifetimes are denoted $\tau^s$, and the offset time
is denoted $t_0^s$. In the case of splitting we find the offset time to be $t_0^s = 500$, except for $Re=1350$, the largest value studied, where $t_0^s = 800$. It should be noted that obtaining splitting times becomes delicate at $Re=1350$ because turbulence spreads in less distinct bands. The survival distributions for various $Re$ are plotted in Fig.~\ref{fig:stat_split}. As with decay, these data are again consistent with exponential distributions.


At $Re=900$ and $Re=1100$, some of the runs reach the cut-off time $t_{\rm f}=10^5$. From a total simulation time of about $10^6$ time units, we registered only 10 decay events at $Re=900$ and 25 splitting events at $Re=1100$, immediately showing that the characteristic lifetimes at these values of $Re$ are on the order of $10^5$ for $Re=900$ and $6\times 10^4$ for $Re=1100$.  Investigations at $Re=950$, 1000 and 1050 were performed, but no events occurred before $10^5$ time units. Due to the high numerical cost of sampling at  
these longer time scales, we did not attempt further investigation between $Re=900$ and $Re=1100$. As a result, we observed no case in which both splitting and decay events occurred at the same Reynolds number, unlike for plane Couette flow \cite{shi} and pipe flow \cite{avila2011onset}.

Figure \ref{fig:stat_threshold} shows the estimated mean lifetimes $\tau^d$ and $\tau^s$ as a function of Reynolds number.  
%
For simplicity, the error bars correspond to confidence intervals for censored data of type II \cite{lawless}.
The decay lifetimes increase rapidly as a function of $Re$, while the splitting times decrease rapidly as a function of $Re$. It is clear from the main semi-log plot that both dependencies are faster than exponential. While it is not possible to determine with certainty the functional form of the dependence on $Re$, the data are consistent with a double-exponential form, as shown in the inset where the double log of the lifetimes are plotted as a function of $Re$. The linear fits indicated in the inset are plotted as dashed curves in the main figure. From these curves one can estimate the crossing point to be 
$Re_{\text{cross}}\simeq 965$ with a corresponding time-scale of about $3 \times 10^6$. The extrapolation of the data means that these values are only approximate. Nevertheless, we can be sure that the timescale of the crossing in our case is significantly above the crossing timescale of about $2 \times 10^4$ found in a similar study of plane Couette flow \cite{shi}, and it appears to be about a factor of 10 less than the value $2 \times 10^7$ found for pipe flow \cite{avila2011onset}.

\section{Discussion and conclusion}
\label{sec:Discussion}

We have studied the behavior of oblique turbulent bands in plane channel flow using narrow tilted computational domains. Bands in such domains have fixed angle with respect to the streamwise direction and are effectively infinitely long, with no large-scale variation along the band. We have measured the propagation velocity of these bands as a function of Reynolds number and inter-band spacing and found that band speed is affected by band spacing at  distances greater than previously assumed \cite{tuckerman2014turbulent}.

After long times, bands either decay to laminar flow or else split into two bands. Survival distributions obtained from many realizations of these events confirm that both processes are effectively memoryless, with characteristic lifetimes $\tau^d(Re)$ and $\tau^s(Re)$, respectively. 
%
%
The dependence of these lifetimes on $Re$ is super-exponential and consistent with a double-exponential scaling. Fitting the data with double-exponential forms, we estimate that the lifetimes cross at $Re_\text{cross} \simeq 965$, at about $3 \times 10^6$ advective time units. Below $Re_\text{cross}$, isolated bands decay at a faster rate than they split, while above $Re_\text{cross}$, isolated bands split at a faster rate than they decay. Hence $Re_\text{cross}$ is very close to the critical point above which turbulence would be sustained in the tilted computational domain. 
Double-exponential scaling is consistent with what has been observed in pipe flow \cite{avila2011onset}. Such scaling is thought to be connected to extreme-value statistics, as first proposed by Goldenfeld \emph{et al.} \cite{goldenfeld2010extreme} and recently examined quantitatively for puff decay in pipe flow by Nemoto \& Alexakis \cite{nemoto2018method,nemoto2020extreme}.

The characteristic times $\tau^d(Re)$ and $\tau^s(Re)$ in plane channel flow are considerably larger than those for plane Couette flow in a similar computational domain by Shi \emph{et al.}~\cite{shi}, who found that splitting and decay lifetimes cross at about $2 \times 10^4$ advective time units. Time scales in plane channel flow are closer to those in pipe flow, where Avila \emph{et al.}~\cite{avila2011onset} found that lifetimes cross at about $2 \times 10^7$ advective time units. 
The higher crossing times in plane channel flow and pipe flow pose a challenge for determining the exact crossing point. A practical consequence of this higher crossing time is that near the crossing Reynolds number, the flow has a greater tendency to appear to be at equilibrium, with neither decay nor splitting events observed over long times.

We also note that turbulent puffs in both pipe flow \cite{barkley2015rise,song2017speed} and channel flow move slightly faster than the bulk flow for low $Re$ and slightly slower for high $Re$; in both flows, the propagation speed becomes equal to $U_{\rm bulk}$ at a Reynolds number close to the critical point. It is possible that an explanation will be found that relates the propagation speed with the critical point.

Our crossover Reynolds number $Re_\text{cross} \simeq 965$ is close to what Shimizu~\& Manneville \cite{ShimizuPRF2019} called a plausible 2D-DP threshold.
These authors carried out channel flow simulations in a large domain and used the 2D-DP power law to extrapolate the turbulent fraction to zero, leading to a threshold of $Re_\text{DP}=905$ or 984, depending on how the pressure-driven Reynolds number is converted to a bulk Reynolds number. (They did not, however, attempt to verify the other critical exponents associated with 2D-DP since they were unable to extend their data sufficiently close to $Re_\text{DP}$; see paragraph below.)
This agreement between the lifetime crossing point obtained in our narrow tilted domain and the transition threshold obtained in the full planar setting for plane channel flow corroborates similar findings for plane Couette flow and stress-free Waleffe flow. Specifically, the decay-splitting lifetime crossing in tilted plane Couette flow was found by Shi \emph{et al.}~\cite{shi} to occur at $Re \simeq 325$. The transition point in the planar case is not known precisely, but it has been estimated by Bottin \emph{et al.}~\cite{bottin1998statistical,bottin1998discontinuous} and Duguet \emph{et al.}~\cite{duguet2010formation} to be close to this value. In a truncated model of Waleffe flow, tilted domain simulations indicate \cite{Chantry_private} that the lifetime crossing point is at $Re_c \simeq 174$. The critical point in a very large domain was computed accurately by Chantry \emph{et al.}~\cite{chantry_universal} to be $Re_c=173.80$.
Heuristically some agreement between the two types of domains could be expected on the grounds that the onset of sustained turbulence is associated with its stabilization in a modified shear profile \cite{barkley2011simplifying,barkley2016theoretical,song2017speed} and a narrow tilted domain quantitatively captures this process. 
Nevertheless, the very close agreement between the thresholds in tilted and planar domains in several flows is not completely understood.

Shimizu~\& Manneville \cite{ShimizuPRF2019} were prevented from approaching their estimate of $Re_\text{DP}$ when lowering $Re$
by a transition to what they called the one-sided regime.
Flows in this regime contain bands of long but finite length which grow via the production of streaks at their stronger downstream heads \cite{xiong2015turbulent,Kanazawa_thesis,tao2018extended,xiao2020growth}. This regime thus shows a strong asymmetry between the upstream and downstream directions and therefore has no counterpart in plane Couette flow; isolated bands in plane Couette flow are transient \cite{manneville2011decay,chantry_universal,lu2019threshold}.
In the one-sided regime, bands eventually all have the same orientation of about $45^\circ$ from the streamwise direction and do not form a regular pattern. 
Since an essential feature of this regime is the long but finite length of the bands, it cannot be simulated using narrow tilted domains. This can be viewed as a shortcoming of the tilted domain in capturing the full dynamics of channel flow, but it also has the advantage of allowing us to study channel flow with the one-sided regime excluded.

We have described the evolution of a band in a narrow tilted domain during a decay or a splitting event via Fourier spectral decomposition. During a band decay, small-scale structures, streaks and rolls, are damped more quickly, increasing the relative prominence of the large-scale flow parallel to  \cite{coles1966progress,barkley2007mean,chantry2016turbulent,ShimizuPRF2019,xiao2020growth} or around \cite{lemoult2014turbulent,ShimizuPRF2019,xiao2020growth,wesfreid2020}
a turbulent patch or band. All of our realizations have the same exponential decay rate at the end of the process.

Fourier analyses show that large-scale spectral components are correlated throughout the life of a band, but undergo opposite trends during a splitting event, due to one- and two-band interactions. By examining several realizations of band splitting, we find that the first three $z$-Fourier modes follow approximately the same path during the transition from one band to two bands. 
This characterization of the splitting pathway resembles transitions in other turbulent fluid systems for which rare-event algorithms have been applied to assess long time scales
associated with infrequent events.
This has been carried out in \cite{bouchet2019rare} for barotropic jet dynamics in the atmosphere and in \cite{rolland2018extremely} 
for a stochastic two-variable model that reproduces transitional turbulence \cite{barkley2016theoretical}. 
We are currently working on applying this strategy to the study of turbulent band splitting.

\begin{acknowledgments}

The calculations for this work were performed using high performance computing resources provided by the Grand Equipement National de Calcul Intensif at the Institut du D\'eveloppement et des Ressources en Informatique Scientifique (IDRIS, CNRS) through grant A0062A01119. This work was supported by a grant from the Simons Foundation (Grant number 662985, NG). We wish to thank Yohann Duguet, Florian Reetz, Alessia Ferraro, Tao Liu, Jose-Eduardo Wesfreid and Beno\^it Semin for helpful discussions.

\end{acknowledgments}




\bibliography{bib}

\end{document}